\documentclass[%
 reprint,
 superscriptaddress,
 amsmath,amssymb,
 aps,
]{revtex4-2}

\usepackage{graphicx}
\usepackage{xfrac}
\usepackage{hyperref}
\hypersetup{colorlinks,urlcolor=blue, linkcolor=blue}
\begin{document}

\title{Chemically Tuning Room Temperature Pulsed Optically Detected Magnetic Resonance}

\author{Sarah K. Mann}
\affiliation{James Watt School of Engineering, University of Glasgow, Glasgow, G12 8QQ, UK.}

\author{Angus Cowley-Semple}
\affiliation{James Watt School of Engineering, University of Glasgow, Glasgow, G12 8QQ, UK.}

\author{Emma Bryan}
\affiliation{Department of Materials and London Centre for Nanotechnology, Imperial College London, Prince Consort Road, London, SW7 2AZ, UK.}

\author{Ziqiu Huang}
\affiliation{Department of Materials and London Centre for Nanotechnology, Imperial College London, Prince Consort Road, London, SW7 2AZ, UK.}

\author{Sandrine Heutz}
\affiliation{Department of Materials and London Centre for Nanotechnology, Imperial College London, Prince Consort Road, London, SW7 2AZ, UK.}

\author{Max Attwood}
\email{m.attwood@imperial.ac.uk}
\affiliation{Department of Materials and London Centre for Nanotechnology, Imperial College London, Prince Consort Road, London, SW7 2AZ, UK.}

\author{Sam L. Bayliss}
\email{sam.bayliss@glasgow.ac.uk}
\affiliation{James Watt School of Engineering, University of Glasgow, Glasgow, G12 8QQ, UK.}

\begin{abstract}
Optical detection of magnetic resonance enables spin-based quantum sensing with high spatial resolution and sensitivity---even at room temperature---as exemplified by solid-state defects. Molecular systems provide a complementary, chemically tunable, platform for room-temperature optically detected magnetic resonance (ODMR)-based quantum sensing. A critical parameter governing sensing sensitivity is the optical contrast---i.e., the difference in emission between two spin states. In state-of-the-art solid-state defects such as the nitrogen-vacancy center in diamond, this contrast is approximately 30\%. Here, capitalizing on chemical tunability, we show that room-temperature ODMR contrasts of 40\% can be achieved in molecules. Using a nitrogen-substituted analogue of pentacene (6,13-diazapentacene), we enhance contrast compared to pentacene and, by determining the triplet kinetics through time-dependent pulsed ODMR, show how this arises from accelerated anisotropic intersystem crossing. Furthermore, we translate high-contrast room-temperature pulsed ODMR to self-assembled nanocrystals. Overall, our findings highlight the synthetic handles available to optically readable molecular spins and the opportunities to capitalize on chemical tunability for room-temperature quantum sensing.
\end{abstract}

\maketitle
\section{Introduction}

Optically addressable spins are emerging as powerful quantum sensors for detecting physical quantities including magnetic and  electric fields, temperature, and strain. A prime example is the nitrogen-vacancy (NV) center in diamond, a solid-state defect that enables room-temperature, nanoscale spin-based sensing \cite{balasubramanian2008nanoscale, taylor2008high, doherty2013nitrogen, schirhagl2014nitrogen}, and has realized remarkable demonstrations including sub-cellular magnetic imaging of living cells \cite{LeSage2013} and wide-field imaging of neuron activity \cite{Hall2012}. While great progress is being made exploring different solid-state defects \cite{Koehl2011, awschalom2018quantum, Gottscholl2020, stern2024quantum}, optically addressable \textit{molecular} spins offer a complementary approach for quantum sensing with their chemical tunability and nanoscale modularity holding promise for tailor-made functionality and target integration \cite{wasielewski2020exploiting, yu2021molecular, atzori2019second, Gaita2019}. A key parameter determining the sensitivity of such quantum sensors is the optical spin contrast, $C$, i.e., the normalized difference in photoluminescence (PL) between two spin states, $\Delta\text{PL}/\text{PL}$, which is typically 30\% for the NV center \cite{schirhagl2014nitrogen}. This parameter is a key target for optimization, since sensing sensitivity is proportional to $\sfrac{1}{C}$ \cite{Barry2020}, and the ability to \textit{synthetically} enhance room-temperature contrast would be a valuable asset for quantum sensing, uniquely possible through a chemical platform.

\begin{figure}[h!tb]
    \centering
    \includegraphics[width=\columnwidth]{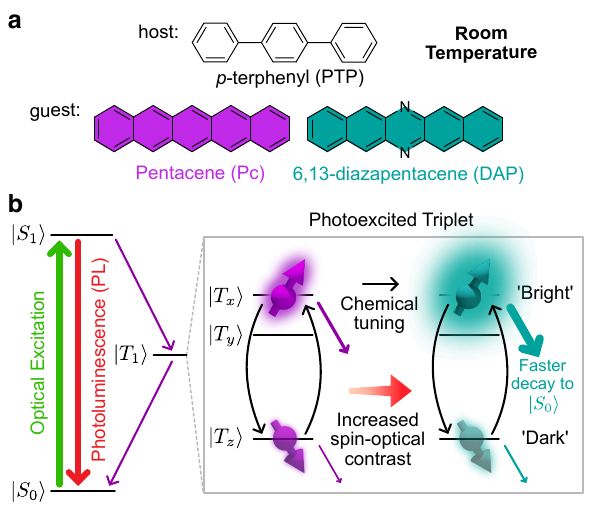} 
    \caption{\textbf{Chemically tuning room-temperature optically detected magnetic resonance (ODMR).} \textbf{a} Chemical structures of host (PTP) and guest (Pc, DAP) molecules. \textbf{b} Simplified energy level diagram illustrating the formation and decay of the photoexcited triplet states in Pc and DAP, along with how ODMR contrast can be enhanced due to modified spin dynamics, producing a greater differentiation between `bright' and `dark' spin sublevels.}
    \label{fig1}
\end{figure}

While recent work has shown promising spin-optical functionality using ground-state molecular spins \cite{bayliss2020optically, serrano2022ultra, gorgon2023reversible, quintes2023properties, kopp2024luminescent, chowdhury2024optical, sutcliffe2024ultrafast, shin2024toward}, molecular photoexcited triplet states in organic chromophores also hold promise for quantum sensing due to their ubiquity, coherence \cite{Sloop1981}, and ability to support optical readout \cite{Wrachtrup1993, Kohler1993, moro2022room}---and in particular, room-temperature pulsed optically detected magnetic resonance (ODMR) \cite{Mena2024, Singh2025}, recently reported for pentacene (Pc) doped in \textit{para}-terphenyl (PTP): see structures in Figure \ref{fig1}a. Demonstrations of room-temperature ODMR in fluorescent proteins \cite{Feder2024, abrahams2024quantum} further highlight the potential of organic photoexcited triplets for quantum sensing. More broadly, the synthetic handles available to molecular quantum sensors offer  rich deployment strategies---e.g., thin films \cite{Mena2024}, spin-labels \cite{valentin2014, hintze2016laser, Bertran2021} and nanoparticles \cite{Pansare2014, Ishiwata2025}---and the atomistic tunability with which to iteratively enhance their properties. Here, using a nitrogen-substituted pentacene, 6,13-diazapentacene (DAP; Figure \ref{fig1}a), we show how room-temperature optical-spin contrast can be chemically enhanced to 40\%. We elucidate the underlying mechanism for this using room-temperature pulsed ODMR to extract the triplet dynamics, and additionally show the opportunities of molecular materials synthesis through high-contrast pulsed ODMR in self-assembled DAP nanocrystals.

DAP:PTP has been a compelling candidate for spin-based quantum technologies, finding application as a maser gain medium \cite{Ng2023}, and dynamic nuclear polarization agent \cite{Kouno2019}. Importantly, nitrogen substitution results in significantly faster excited-state dynamics compared to Pc:PTP \cite{Ng2023,Kouno2019}. Since efficient ODMR relies on distinguishing spin sublevels through their kinetics, while out-competing spin-lattice relaxation, fast (and anisotropic) intersystem crossing (ISC) from triplet sublevels to the ground state indicate promise for enhancing ODMR contrast.

To optimize optical-spin contrast, our aim is to enhance the effective difference in brightness of two triplet sublevels. Figure \ref{fig1}b shows an energy-level diagram for Pc/DAP illustrating the key processes involved in the formation and decay of the photoexcited triplet state. Upon optical excitation, molecules are promoted from the singlet ground state, $|S_0 \rangle$, to the singlet excited state, $|S_1 \rangle$. From there, they can either return to $|S_0 \rangle$, emitting PL, or undergo spin-selective ISC to populate the triplet state, $|T_1 \rangle$ (with an overall  yield of $\simeq$65\% \cite{Takeda2002, Ng2023}). Zero-field splitting lifts the degeneracy of the triplet sublevels in the absence of a magnetic field, resulting in three distinct sublevels: $|T_x\rangle$, $|T_y\rangle$, and $|T_z\rangle$ ($x$, $y$, and $z$ axes correspond to the molecule's long, short, and out-of-plane directions, respectively.) Once in the triplet state, the initial spin polarization, rates of triplet depopulation, and spin-lattice relaxation determine the optical contrast. Faster decay from a triplet sublevel to $|S_0 \rangle$ allows for more rapid re-excitation and PL emission, resulting in a `bright' sublevel, whereas slower decay leads to a `dark' sublevel (illustrated in Figure \ref{fig1}b). (For clarity, we note that we use \textit{fluorescence} of the $|S_1 \rangle\rightarrow|S_0 \rangle$ transition for spin readout, rather than phosphorescence from the $|T_1 \rangle\rightarrow|S_0 \rangle$ transition.) This spin-dependent brightness enables spin-state readout via ODMR: when microwaves are applied matching the triplet sublevels’ transition frequencies, populations are redistributed between the bright and dark sublevels, leading to changes in PL intensity \cite{Kohler1993, Wrachtrup1993}. This mechanism highlights the potential for tailoring spin dynamics to enhance optical spin contrast as we show below.

\section{Results and Discussion}

\begin{figure*}[h!tb]
    \centering
    \includegraphics[width=\textwidth]{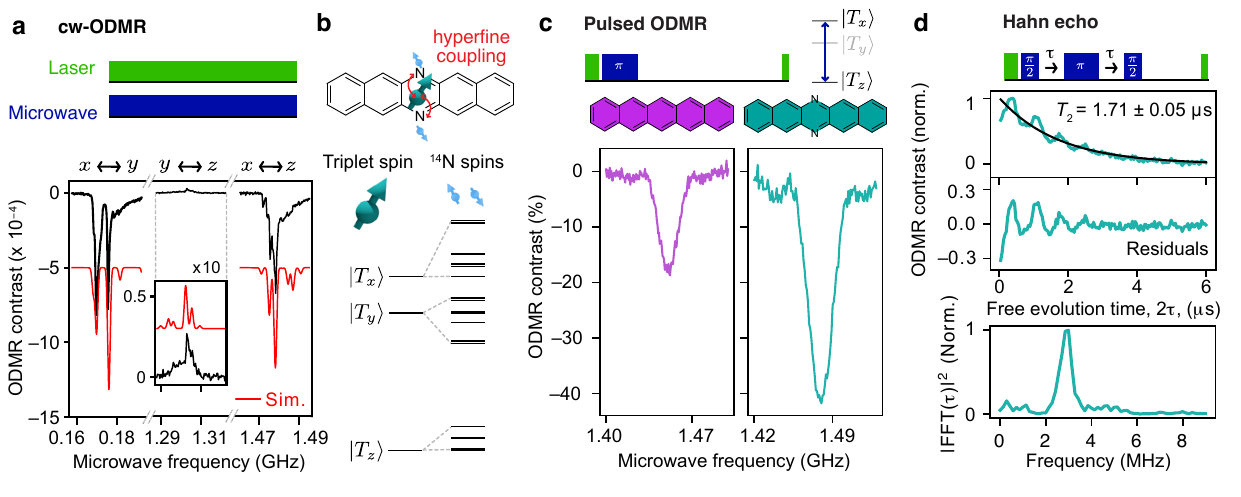} 
    \caption{\textbf{40\% room-temperature optically detected magnetic resonance contrast in a molecular system.} All experiments were performed at zero applied magnetic field. \textbf{a} Continuous-wave ODMR spectrum of a DAP:PTP single crystal (black) compared with EasySpin simulations (red) using DFT-calculated hyperfine/quadrupole parameters for the \textsuperscript{14}N spins. \textbf{b} Illustration of the hyperfine coupling to the \textsuperscript{14}N nuclei. \textbf{c} Pulsed ODMR ($|T_x\rangle \leftrightarrow |T_z\rangle$ transition) spectra of Pc:PTP (0.1\% doped) and DAP:PTP (0.5\% doped) 100 nm thin films showing 40\% contrast in DAP. \textbf{d} Optically detected Hahn-echo of a DAP:PTP single crystal, yielding $T_2 = 1.71 \pm 0.05$ $\mu$s, determined from an exponential fit (black line), along with electron spin-echo envelope modulation oscillations determined by subtracting the exponential fit (middle) and Fourier-transforming (bottom).}
    \noindent\rule{\textwidth}{0.5pt}
    \label{fig2}
\end{figure*}

\textit{Room-temperature optically detected magnetic resonance of diazapentacene}.---To demonstrate the effect of nitrogen substitution on room-temperature pulsed ODMR contrast, we first measure the continuous-wave (cw) ODMR of a single crystal of DAP doped at 0.01\% in PTP (Figure \ref{fig2}a). In comparison to Pc, which shows a single peak for each triplet transition \cite{Mena2024}, DAP shows additional splittings in the ODMR spectrum due to coupling to the two \textsuperscript{14}N spins (with $I=1$; Figure \ref{fig2}b). The \textsuperscript{14}N hyperfine couplings are larger than those of protons due to the greater electron spin density on the nitrogen. The ODMR spectrum shows close agreement with simulations using density functional theory (DFT) calculated \textsuperscript{14}N hyperfine and quadrupole interactions (red solid line; see the Supporting Information for details), where we calculate diagonal hyperfine and quadrupole matrices with components [$A_{xx}$, $A_{yy}$, $A_{zz}$] = [$-0.79$, $-0.99$, $23$] MHz and [$Q_{xx}$, $Q_{yy}$, $Q_{zz}$] = [$0.99$, $-2.2$, $1.2$] MHz (aligned with the zero-field splitting tensor). The best-fit zero-field splitting parameters are $D = 1390.5$ and $E = -84.9$ MHz. The negative ODMR contrast of the $|T_x\rangle \leftrightarrow |T_y\rangle$ and $|T_x\rangle \leftrightarrow |T_z\rangle$ transitions, and the weaker positive contrast of the $|T_y\rangle \leftrightarrow |T_z\rangle$ transition is similar to pentacene \cite{Mena2024}.
 
\textit{Enhanced optical-spin contrast in diazapentacene}.---We next demonstrate room-temperature optically-detected coherent control of DAP (the second molecular system to demonstrate such behavior after Pc:PTP \cite{Mena2024, Singh2025}). For maximal contrast, we use thin films which facilitate more efficient excitation of the population from the ground state compared to bulk crystals. We observe a significantly higher pulsed ODMR contrast of 40\% in DAP compared to Pc (18\%) under optimized conditions in 100 nm doped PTP thin films (Figure \ref{fig2}c). This 40\% contrast exceeds the typical 30\% contrast found for nitrogen-vacancy centres under single-spin conditions \cite{schirhagl2014nitrogen, Wirtitsch2023}, with potential for further improvements through molecular control. Since sensing sensitivity scales (approximately) as \(\eta^V\propto\frac{1}{C\sqrt{n_{\text{avg}}c_{s}}}\frac{\sqrt{t_{\text{overhead}}}}{T_{2}^{\chi}}\) \cite{Barry2020}---where $C$ is the optical-spin contrast, $n_{\text{avg}}$ is the average number of photons collected per spin per readout, $c_s$ is the spin density, $t_{\text{overhead}}$ the measurement overhead time, and $T_{2}^{\chi}$ is $T_2$ (or the equivalent under dynamical decoupling) for AC sensing and $T_2^\star$ for DC sensing---and is therefore inversely proportional to the optical-spin contrast, $C$, optimizing this parameter is key (particularly as other parameters, such as spin concentration, feature as a square-root, reducing their impact).

To determine that DAP's room-temperature coherence time ($T_2$) is not adversely affected by nitrogen substitution, we use an optically detected Hahn-echo sequence to extract $T_2 = 1.71 \pm 0.05$ $\mu$s (Figure \ref{fig2}d), comparable to Pc \cite{Mena2024} (due to the similar \textsuperscript{1}H-dominated nuclear spin bath). The Hahn-echo trace exhibits oscillatory behaviour, i.e., electron spin echo envelope modulation (ESEEM) \cite{Mims1972}, which results from coherent coupling of the triplet spin with the \textsuperscript{14}N nuclei. Notably, this ESEEM is observed at zero magnetic field \cite{Van1979, Janes1983, Weis1998} (in contrast to demonstrations using electron paramagnetic resonance spectroscopy under an applied field). We Fourier-transform the residual oscillations---obtained by subtracting the exponential decay fit (Figure \ref{fig2}d, centre)---to extract the ESEEM frequency spectrum (Figure \ref{fig2}d, bottom), with the 3 MHz oscillation frequency agreeing with the nuclear quadrupole transition frequency $Q_{xx}-Q_{yy}$ determined from our DFT calculations for the \textsuperscript{14}N nuclei. These strongly coupled nuclear spins  (hyperfine coupling greater than electron-spin line-widths), provide a future resource for enhanced quantum sensing through electron-nuclear registers, exemplified by demonstrations with the NV center \cite{zaiser2016enhancing, Arunkumar2023}.

\begin{figure}[h!tb]
    \centering
    \includegraphics[width=\columnwidth]{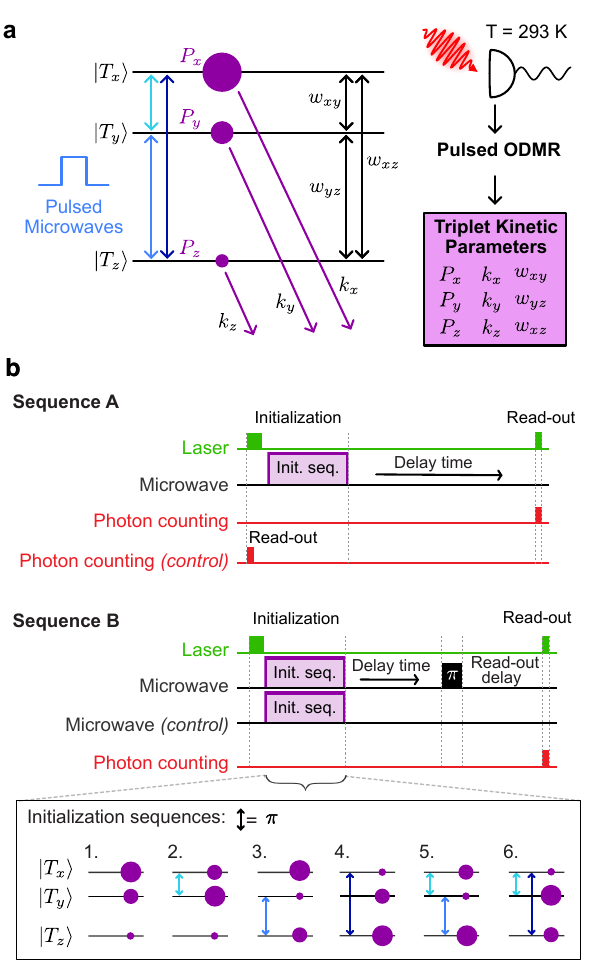} 
    \caption{\textbf{Quantifying triplet spin dynamics through room-temperature pulsed ODMR.} \textbf{a} Energy-level diagram illustrating the processes involved in the formation and decay of the photoexcited triplet state. Purple circles represent the relative sublevel populations. Through room-temperature pulsed ODMR, we obtain the parameters describing the triplet spin dynamics. \textbf{b} Pulse sequences A and B used to determine the dynamics. The triplet sublevels are prepared in six different states (initialization sequences 1-6) via microwave $\pi$ pulses on the three spin transitions.}
    \label{fig3}
\end{figure}

\textit{Quantifying spin dynamics}.--- A key question arises: why is the ODMR contrast enhanced in DAP compared to Pc? To investigate this, we designed a series of pulsed ODMR experiments to gain detailed information about the underlying dynamics. Figure \ref{fig3}a shows the parameters defining the triplet spin dynamics. The formation of the triplet state via ISC is highly spin selective, resulting in triplet sublevel populations $P_x:P_y:P_z$ (previously reported as $\simeq 0.76:0.16:0.08$ for Pc \cite{Sloop1981} and $\simeq 0.60:0.21:0.19$ for DAP \cite{Ng2023}). Triplet sublevels decay anisotropically back to the singlet ground state at rates $k_i$ ($i = x, y, z$) and spin-lattice relaxation transfers populations between the sublevels at rates $w_{ij}$ ($i \neq j$). Quantifying triplet spin dynamics can therefore be challenging as it depends on nine parameters: the triplet decay rates ($k_x$, $k_y$, $k_z$), the spin-lattice relaxation rates ($w_{xy}$, $w_{yz}$, $w_{xz}$), and the initial populations of the triplet sub-levels following laser excitation ($P_x$, $P_y$, $P_z$). These parameters can be difficult to quantify with other techniques---such as transient electron paramagnetic resonance (Tr-EPR) spectroscopy---due to the challenge of tuning EPR resonators across all transition frequencies \cite{Wu2019, Ng2023, Attwood2023}. To sensitively determine the triplet kinetic parameters, here we capitalize on the opportunities of room-temperature pulsed ODMR. This approach enables us to deploy broadband pulsed microwave control to realize a large number of distinct experiments---e.g., shuffling initial triplet populations to prepare six initial conditions---while using the sensitivity of ODMR contrast to triplet kinetics to determine the key parameters outlined above. 

We perform a total of 22 different measurements using two distinct sequences, A \& B (Figure \ref{fig3}b) which we apply to single-crystal samples (to minimize inhomogeneity arising from variations in molecular orientation). Each sequence starts with the triplet initialized in one of six different states which we can prepare through the application of microwave $\pi$ pulses on different transitions (initialization sequences 1--6) to shuffle the initial populations (Figure \ref{fig3}b). In Sequence A, following initialization, we vary the delay time before using the PL from a short laser read-out pulse to probe the repopulation of the ground state, $|S_0\rangle$. This signal is normalized to the PL arising from the unperturbed ground-state population (control sequence), resulting in a signal that is a direct measure of the $|S_0\rangle$ population. In Sequence B, we use a similar experiment, but rather than probing the ground-state population directly following the delay time, we probe the effect of a microwave $\pi$ pulse resonant with one of the three spin transitions. A delay time (on the order of microseconds) following the microwave inversion pulse allows repopulation of the ground state in a spin-dependent fashion, which we probe through the PL from a final read-out laser pulse. This signal is referenced to the PL obtained from the control sequence without the final $\pi$-pulse. The combination of the six initialization sequences with Sequence A gives six experiments, while the combination of the six initialization sequences and three choices for the delayed $\pi$-pulse ($|T_x\rangle \leftrightarrow |T_y\rangle$, $|T_y\rangle \leftrightarrow |T_z\rangle$ and $|T_x\rangle \leftrightarrow |T_z\rangle$) in sequence B, gives a further 18 measurements, and therefore a total of 24 possible time-dependent measurements. For convenience, we exclude two of these which require control of all three microwave frequencies, leaving 22 distinct measurements.

\begin{figure}
    \centering
    \includegraphics[width=\columnwidth]{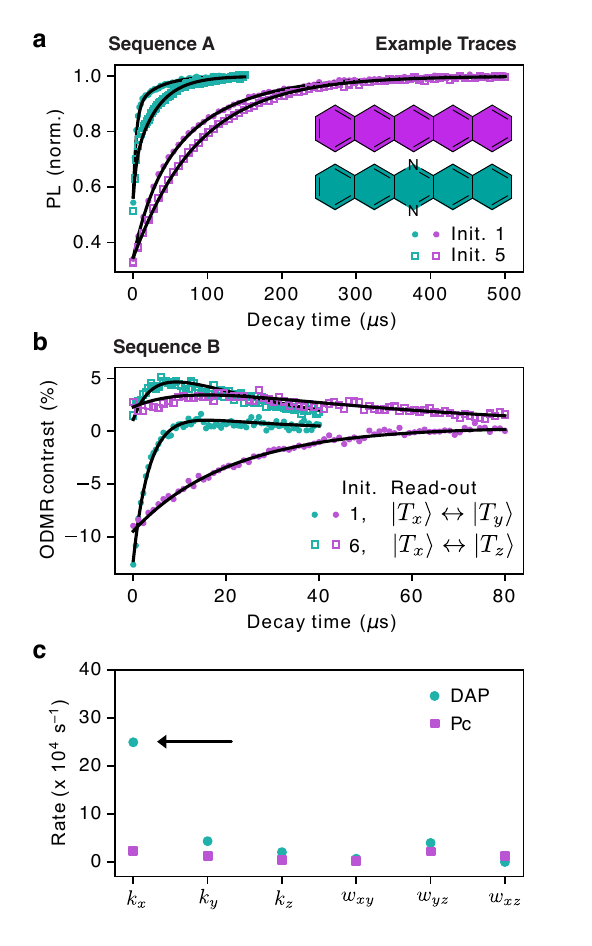}
    \caption{\textbf{Triplet spin dynamics from optically detected relaxation measurements.} Example relaxation curves for DAP:PTP and Pc:PTP (0.01\% single crystals) recorded using \textbf{a} sequence A, and  \textbf{b} sequence B, along with global fits (black lines). \textbf{c} Best-fit values of the spin dynamics rates for DAP:PTP and Pc:PTP.}
    \label{fig4}
\end{figure}

\textit{Benchmarking dynamics extraction through pulsed ODMR}.---Before applying this technique to DAP:PTP, we first benchmark it using Pc:PTP (0.01\% doped single crystal), whose rates have previously been characterized at zero-field and room temperature \cite{Wu2019, Wu2020}, demonstrating reduced uncertainties compared to previous measurements. Example curves for Sequence A and B are shown in Figures \ref{fig4}a and \ref{fig4}b, respectively, with all 22 curves for Pc:PTP shown in the Supporting Information (Figure S4). We globally fit all relaxation measurements to extract the 9 parameters (see Supporting Information for fitting details). The resulting parameters ($k_{i}, w_{ij}, P_i$) are shown in Figure \ref{fig4}c and Tables S1-2 and are in close agreement with those previously reported (Tables S1-S2) \cite{Wu2019, Wu2020}. Importantly, our approach yields lower fitting errors, which we attribute to the increased amount of information extracted from this pulsed ODMR approach, demonstrating this method's potential to characterize room-temperature spin dynamics with high precision. Crucially, this method opens up sensitive characterization on less well-studied systems, which we now realize on DAP.

\textit{Tuned spin dynamics: diazapentacene}.---We extend our pulsed-ODMR method for characterizing spin dynamics to DAP:PTP (0.01\% single crystal). Figures \ref{fig4}a and \ref{fig4}b show example relaxation curves while all 22 curves and fits are shown in Figure S5. We are able to unambiguously extract a full set of parameters (Figure \ref{fig4}c, Tables S1-S2), something which is challenging with other approaches \cite{Ng2023}. The curves show that the spin dynamics of DAP are markedly different to Pc, with significantly faster triplet depopulation (Figures \ref{fig4}a and \ref{fig4}b). Strikingly, the depopulation rate, $k_x = (24.9 \pm 0.2)\times10^4\,\mathrm{s}^{-1}$ (decay time of 4.0 $\mu$s; marked by the arrow in Figure \ref{fig4}c) is approximately 10-times faster than for Pc. In addition,  $k_y$ and $k_z$ are approximately four-times faster than in Pc, while following the same trend of $k_x > k_y > k_z$. (These results are in agreement with average triplet lifetimes of 3.3 $\mu$s, obtained by transient absorption \cite{Kouno2019}, and 4.6 $\mu$s, determined by EPR spectroscopy \cite{Ng2023}.) 

These kinetic parameters highlight two factors in particular which account for the higher ODMR contrast in DAP compared to Pc. Firstly, the ratio of $k_x/k_z = 12$ is higher for DAP compared to the $k_x/k_z = 5$ we extract for Pc. This larger difference in depopulation rates between the $|T_x\rangle$ and $|T_z\rangle$ sublevels enables a greater imbalance in ground-state repopulation---and therefore ability to emit PL under re-excitation---thereby enhancing the ODMR contrast. Secondly, triplet depopulation occurs on a faster timescale relative to spin-lattice relaxation for DAP compared to Pc. This reduces the mixing of triplet sublevel populations before they decay, thereby improving spin-state readout. Importantly, since the spin-lattice relaxation rates in DAP are not enhanced to the same extent as the depopulation rates, our results demonstrate that high-contrast room-temperature pulsed-ODMR with molecules does not necessarily require improvements in spin-lattice relaxation.

The faster dynamics of DAP compared to Pc are further beneficial for ODMR measurements as a long triplet lifetime can otherwise provide a bottleneck to the experimental repetition rate (since the population needs to return to the ground state before restarting a measurement). The 10-times faster decay of $|T_x\rangle$ in DAP compared to Pc enables a higher repetition rate, and more photons to be collected per unit time due to faster cycling. Finally, we note that fast decay of $|T_x\rangle$ need not limit the available spin manipulation time: the shorter-lived $|T_x\rangle$ population can be transferred (via a microwave pulse) to the longer-lived $|T_y\rangle$ / $|T_z\rangle$ sublevels (which can serve as the qubit), with population transferred back to $|T_x\rangle$ for effective readout \cite{Mena2024}.

\textit{Physical origin of the modified dynamics}.---Having determined how the modified spin dynamics under nitrogen substitution lead to increased contrast, we now turn to their physical origin. In planar aromatic molecules, triplet depopulation rates are dominated by non-radiative transitions driven by vibronic spin-orbit coupling \cite{Siebrand1970, Metz1973, Clarke1976}. Such a vibration-mediated mechanism is required to mix $\pi\pi^*$-states with states with $\sigma-$ or $n$-type character, since direct spin-orbit coupling between $\pi\pi^*$-states is weak (as described by El-Sayed's rule \cite{el1963spin, marian2021understanding}). The different wavefunctions of each triplet sublevel give rise to distinct spin-orbit interactions, leading to  $k_x > k_y > k_z$ in Pc and related molecules \cite{Gromer1972, Metz1973, Clarke1976, Wu2019}. For nitrogen-containing heterocycles, such as DAP, the nitrogen lone pair introduces new low-energy $n\pi^*$-states which can more effectively promote mixing with $\pi\pi^*$-states, thereby accelerating ISC \cite{Antheunis1974}. Furthermore, for the nitrogen lone-pair parallel to the molecular y-axis---as in DAP---the increased spin-orbit interaction is most significant for $|T_x\rangle$, thereby most prominently enhancing $k_x$. Our observations are in agreement with Antheunis et al. \cite{Antheunis1974}, who showed that in going from anthracene to its nitrogen-substituted derivatives, acridine and phenazine, the largest increase in depopulation rate is for $k_x$. Interestingly, we do not observe a similar anisotropic enhancement in $P_x$ in going from Pc to DAP. We assign this to the different states involved in triplet population ($|S_1\rangle \rightarrow |T_2\rangle$, where  $|T_2\rangle$ is the second excited triplet) compared to depopulation ($|T_1\rangle \rightarrow |S_0\rangle$) meaning the nitrogen lone pair can contribute in distinct ways to these processes.

\begin{figure}[htb]
    \centering
    \includegraphics[width=\columnwidth]{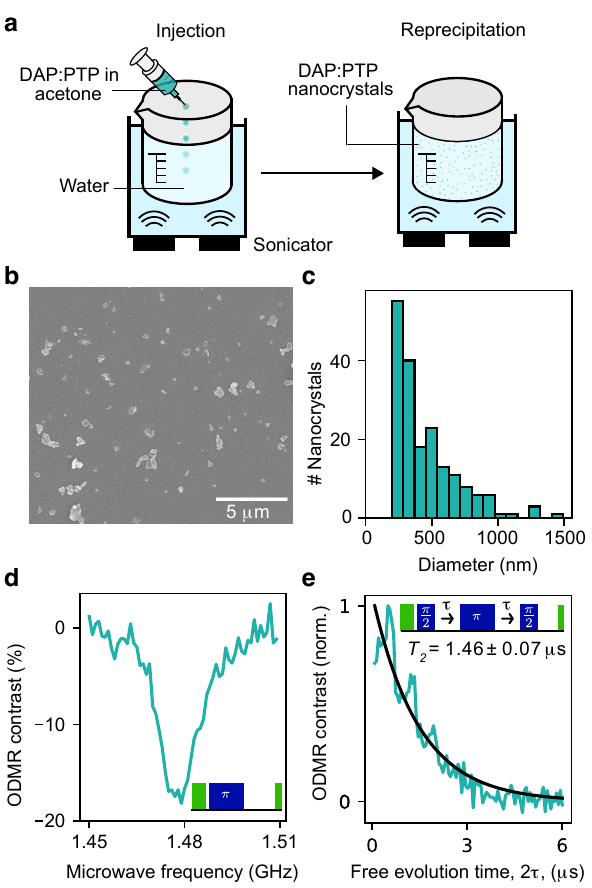}
    \caption{\textbf{Room-temperature pulsed optically detected magnetic resonance of self-assembled DAP:PTP nanocrystals}. \textbf{a} Schematic of nanocrystal growth through the reprecipitation method. \textbf{b} Scanning electron microscope (SEM) image of DAP:PTP nanocrystals. \textbf{c} Nanocrystal size distribution extracted from the SEM data. \textbf{d} Pulsed-ODMR spectrum showing 18\% contrast along with the pulse sequence (inset). \textbf{e} Hahn echo along with an exponential fit (black line) yielding $T_2=1.46\pm0.07\,\mu$s. The inset shows the pulse sequence.}
    \label{fig5}
\end{figure}

\textit{Pulsed ODMR of self-assembled DAP:PTP nanocrystals under ambient conditions:}---Finally, we extend room-temperature optically detected coherent control to self-assembled nanocrystals of DAP:PTP ($0.1\%$ doping concentration), complementing recent demonstrations of ODMR in ball-milled Pc:PTP nanocrystals \cite{Ishiwata2025}. Nanocrystals are attractive for integration with sensing targets and devices \cite{McGuinness2011, kucsko2013nanometre, Pazzagli2018, toninelli2021single}, while retaining the beneficial spin-optical dynamics of a crystalline environment (e.g., high contrast). Self-assembled DAP:PTP nanocrystals were grown using the solution-based reprecipitation method \cite{Kasai1992, kang2004novel} (Figure \ref{fig5}a), facilitated by the more favorable solubility of DAP compared to Pc. Droplets of DAP:PTP in acetone were injected into a beaker of sonicated water. Gradually, the acetone dissolves in the water, increasing the DAP:PTP concentration in the droplets until they form nanocrystals (whose size and morphology can be controlled by altering thermodynamic conditions and mixing speed \cite{Baba2011, Pazzagli2018}). The nanocrystal solution was filtered to select particles $\lesssim450$ nm and drop-cast onto a silicon substrate for subsequent measurements. Figure \ref{fig5}b shows a scanning electron microscope (SEM) image of the nanocrystals which we use to determine their size distribution (Figure \ref{fig5}c), finding a 447 nm mean diameter (see the Supplementary Information for SEM details). The nanocrystal PL spectrum is similar to bulk DAP:PTP (Figure S2), showing the retention of crystalline properties. Pulsed ODMR (Figure \ref{fig5}d), shows 18\% optical contrast and $T_2=1.46\pm0.07\mu$s (Figure \ref{fig5}e; with the same dominant ESEEM frequency as in Figure \ref{fig2}d---see Figure S7 for comparison), demonstrating the preservation of favorable room-temperature ODMR properties in these self-assembled nanocrystals.

\section{Conclusions}
Our work demonstrates the potential of chemical tunability to enhance room-temperature quantum sensing metrics. By minor chemical modifications, we significantly improved room-temperature optical-spin contrast---one of the key parameters influencing sensing sensitivity---to 40\% (which exceeds the typical contrasts of nitrogen vacancy centers in diamond). This increased contrast arises from accelerated anisotropic intersystem facilitated by the lone pair of the substituted nitrogens, highlighting the potential for future synthetic enhancements through control over intersystem crossing dynamics. Our demonstration of characterizing room-temperature photoexcited triplet state dynamics through information-rich pulsed ODMR techniques offers benefit for wider application areas including triplet-based dynamic nuclear polarization \cite{Kouno2019}, and masing \cite{Oxborrow2012}. The room-temperature optically-detected coherent coupling between electrons and nuclei we observe here paves the way for future experiments directly capitalizing on molecules with optically readable strongly coupled electron-nuclear registers \cite{filidou2012ultrafast, Lee2013_ancilla, Arunkumar2023}, and our demonstration of high-contrast spin readout in self-assembled nanocrystals highlights the potential of chemical techniques to synthesize deployable quantum sensors at scale. Overall, our work showcases the promise of a synthetically tunable platform for spin-based quantum sensing that can be iteratively enhanced through chemistry.
 
\subsection*{Supporting Information}
Supporting Information is available at [URL to be added in proof]: experimental details including diagram of the setup, PL spectra, Rabi oscillations, simulation details and fitting procedures, relaxation curves, tabulated fit parameters, additional discussion on spin dynamics, ESEEM of DAP nanocrystals, and powder X-ray diffraction of DAP:PTP films and nanocrystals.

\subsection*{Author contributions}
S.K.M. designed the triplet spin dynamics experiments, performed ODMR measurements, analyzed the ODMR data and performed DFT calculations. A.C. performed nanocrystal ODMR measurements. M.A. prepared and characterized the crystal, film, and nanocrystal samples. E.B. and Z.H. prepared and characterized the thin film and nanocrystal samples. S.H., M.A., and S.L.B. provided oversight and supervision. S.K.M., A.C., M.A., and S.L.B wrote the manuscript with input from all authors.

\subsection*{Acknowledgments}
We thank C. Paunica, A. Mena, and M. Oxborrow for helpful discussions and input, and C. Bonato and H. Stern for helpful feedback on the manuscript. Z. H. and M. A. thank Dr. Ecaterina Ware for assistance with SEM imaging. This work was supported by UK Research and Innovation [grant number MR/W006928/1] and the UK Engineering and Physical Sciences Research Council [grant numbers EP/W027542/1 and EP/V048430/1]. E. B. was supported through the EPSRC and SFI Centre for Doctoral Training in the Advanced Characterization of Materials (CDT-ACM) [Grant No. EP/S023259/1].
 
\subsection*{Competing interests}
The authors declare no competing interests.

\subsection*{Data availability}
The data underlying this work are available at [URL to be added in proof].


\begin{thebibliography}{66}%
\makeatletter
\providecommand \@ifxundefined [1]{%
 \@ifx{#1\undefined}
}%
\providecommand \@ifnum [1]{%
 \ifnum #1\expandafter \@firstoftwo
 \else \expandafter \@secondoftwo
 \fi
}%
\providecommand \@ifx [1]{%
 \ifx #1\expandafter \@firstoftwo
 \else \expandafter \@secondoftwo
 \fi
}%
\providecommand \natexlab [1]{#1}%
\providecommand \enquote  [1]{``#1''}%
\providecommand \bibnamefont  [1]{#1}%
\providecommand \bibfnamefont [1]{#1}%
\providecommand \citenamefont [1]{#1}%
\providecommand \href@noop [0]{\@secondoftwo}%
\providecommand \href [0]{\begingroup \@sanitize@url \@href}%
\providecommand \@href[1]{\@@startlink{#1}\@@href}%
\providecommand \@@href[1]{\endgroup#1\@@endlink}%
\providecommand \@sanitize@url [0]{\catcode `\\12\catcode `\$12\catcode
  `\&12\catcode `\#12\catcode `\^12\catcode `\_12\catcode `\%12\relax}%
\providecommand \@@startlink[1]{}%
\providecommand \@@endlink[0]{}%
\providecommand \url  [0]{\begingroup\@sanitize@url \@url }%
\providecommand \@url [1]{\endgroup\@href {#1}{\urlprefix }}%
\providecommand \urlprefix  [0]{URL }%
\providecommand \Eprint [0]{\href }%
\providecommand \doibase [0]{https://doi.org/}%
\providecommand \selectlanguage [0]{\@gobble}%
\providecommand \bibinfo  [0]{\@secondoftwo}%
\providecommand \bibfield  [0]{\@secondoftwo}%
\providecommand \translation [1]{[#1]}%
\providecommand \BibitemOpen [0]{}%
\providecommand \bibitemStop [0]{}%
\providecommand \bibitemNoStop [0]{.\EOS\space}%
\providecommand \EOS [0]{\spacefactor3000\relax}%
\providecommand \BibitemShut  [1]{\csname bibitem#1\endcsname}%
\let\auto@bib@innerbib\@empty
\bibitem [{\citenamefont {Balasubramanian}\ \emph {et~al.}(2008)\citenamefont
  {Balasubramanian}, \citenamefont {Chan}, \citenamefont {Kolesov},
  \citenamefont {Al-Hmoud}, \citenamefont {Tisler}, \citenamefont {Shin},
  \citenamefont {Kim}, \citenamefont {Wojcik}, \citenamefont {Hemmer},
  \citenamefont {Krueger} \emph {et~al.}}]{balasubramanian2008nanoscale}%
  \BibitemOpen
  \bibfield  {author} {\bibinfo {author} {\bibfnamefont {G.}~\bibnamefont
  {Balasubramanian}}, \bibinfo {author} {\bibfnamefont {I.}~\bibnamefont
  {Chan}}, \bibinfo {author} {\bibfnamefont {R.}~\bibnamefont {Kolesov}},
  \bibinfo {author} {\bibfnamefont {M.}~\bibnamefont {Al-Hmoud}}, \bibinfo
  {author} {\bibfnamefont {J.}~\bibnamefont {Tisler}}, \bibinfo {author}
  {\bibfnamefont {C.}~\bibnamefont {Shin}}, \bibinfo {author} {\bibfnamefont
  {C.}~\bibnamefont {Kim}}, \bibinfo {author} {\bibfnamefont {A.}~\bibnamefont
  {Wojcik}}, \bibinfo {author} {\bibfnamefont {P.~R.}\ \bibnamefont {Hemmer}},
  \bibinfo {author} {\bibfnamefont {A.}~\bibnamefont {Krueger}}, \emph
  {et~al.},\ }\bibfield  {title} {\bibinfo {title} {Nanoscale imaging
  magnetometry with diamond spins under ambient conditions},\ }\href
  {https://doi.org/10.1038/nature07278} {\bibfield  {journal} {\bibinfo
  {journal} {Nature}\ }\textbf {\bibinfo {volume} {455}},\ \bibinfo {pages}
  {648} (\bibinfo {year} {2008})}\BibitemShut {NoStop}%
\bibitem [{\citenamefont {Taylor}\ \emph {et~al.}(2008)\citenamefont {Taylor},
  \citenamefont {Cappellaro}, \citenamefont {Childress}, \citenamefont {Jiang},
  \citenamefont {Budker}, \citenamefont {Hemmer}, \citenamefont {Yacoby},
  \citenamefont {Walsworth},\ and\ \citenamefont {Lukin}}]{taylor2008high}%
  \BibitemOpen
  \bibfield  {author} {\bibinfo {author} {\bibfnamefont {J.~M.}\ \bibnamefont
  {Taylor}}, \bibinfo {author} {\bibfnamefont {P.}~\bibnamefont {Cappellaro}},
  \bibinfo {author} {\bibfnamefont {L.}~\bibnamefont {Childress}}, \bibinfo
  {author} {\bibfnamefont {L.}~\bibnamefont {Jiang}}, \bibinfo {author}
  {\bibfnamefont {D.}~\bibnamefont {Budker}}, \bibinfo {author} {\bibfnamefont
  {P.~R.}\ \bibnamefont {Hemmer}}, \bibinfo {author} {\bibfnamefont
  {A.}~\bibnamefont {Yacoby}}, \bibinfo {author} {\bibfnamefont
  {R.}~\bibnamefont {Walsworth}},\ and\ \bibinfo {author} {\bibfnamefont
  {M.~D.}\ \bibnamefont {Lukin}},\ }\bibfield  {title} {\bibinfo {title}
  {High-sensitivity diamond magnetometer with nanoscale resolution},\ }\href
  {https://doi.org/10.1038/nphys1075} {\bibfield  {journal} {\bibinfo
  {journal} {Nature Physics}\ }\textbf {\bibinfo {volume} {4}},\ \bibinfo
  {pages} {810} (\bibinfo {year} {2008})}\BibitemShut {NoStop}%
\bibitem [{\citenamefont {Doherty}\ \emph {et~al.}(2013)\citenamefont
  {Doherty}, \citenamefont {Manson}, \citenamefont {Delaney}, \citenamefont
  {Jelezko}, \citenamefont {Wrachtrup},\ and\ \citenamefont
  {Hollenberg}}]{doherty2013nitrogen}%
  \BibitemOpen
  \bibfield  {author} {\bibinfo {author} {\bibfnamefont {M.~W.}\ \bibnamefont
  {Doherty}}, \bibinfo {author} {\bibfnamefont {N.~B.}\ \bibnamefont {Manson}},
  \bibinfo {author} {\bibfnamefont {P.}~\bibnamefont {Delaney}}, \bibinfo
  {author} {\bibfnamefont {F.}~\bibnamefont {Jelezko}}, \bibinfo {author}
  {\bibfnamefont {J.}~\bibnamefont {Wrachtrup}},\ and\ \bibinfo {author}
  {\bibfnamefont {L.~C.}\ \bibnamefont {Hollenberg}},\ }\bibfield  {title}
  {\bibinfo {title} {The nitrogen-vacancy colour centre in diamond},\ }\href
  {https://doi.org/10.1016/j.physrep.2013.02.001} {\bibfield  {journal}
  {\bibinfo  {journal} {Physics Reports}\ }\textbf {\bibinfo {volume} {528}},\
  \bibinfo {pages} {1} (\bibinfo {year} {2013})}\BibitemShut {NoStop}%
\bibitem [{\citenamefont {Schirhagl}\ \emph {et~al.}(2014)\citenamefont
  {Schirhagl}, \citenamefont {Chang}, \citenamefont {Loretz},\ and\
  \citenamefont {Degen}}]{schirhagl2014nitrogen}%
  \BibitemOpen
  \bibfield  {author} {\bibinfo {author} {\bibfnamefont {R.}~\bibnamefont
  {Schirhagl}}, \bibinfo {author} {\bibfnamefont {K.}~\bibnamefont {Chang}},
  \bibinfo {author} {\bibfnamefont {M.}~\bibnamefont {Loretz}},\ and\ \bibinfo
  {author} {\bibfnamefont {C.~L.}\ \bibnamefont {Degen}},\ }\bibfield  {title}
  {\bibinfo {title} {Nitrogen-vacancy centers in diamond: Nanoscale sensors for
  physics and biology},\ }\href
  {https://doi.org/10.1146/annurev-physchem-040513-103659} {\bibfield
  {journal} {\bibinfo  {journal} {Annual Review of Physical Chemistry}\
  }\textbf {\bibinfo {volume} {65}},\ \bibinfo {pages} {83} (\bibinfo {year}
  {2014})}\BibitemShut {NoStop}%
\bibitem [{\citenamefont {Le~Sage}\ \emph {et~al.}(2013)\citenamefont
  {Le~Sage}, \citenamefont {Arai}, \citenamefont {Glenn}, \citenamefont
  {DeVience}, \citenamefont {Pham}, \citenamefont {Rahn-Lee}, \citenamefont
  {Lukin}, \citenamefont {Yacoby}, \citenamefont {Komeili},\ and\ \citenamefont
  {Walsworth}}]{LeSage2013}%
  \BibitemOpen
  \bibfield  {author} {\bibinfo {author} {\bibfnamefont {D.}~\bibnamefont
  {Le~Sage}}, \bibinfo {author} {\bibfnamefont {K.}~\bibnamefont {Arai}},
  \bibinfo {author} {\bibfnamefont {D.~R.}\ \bibnamefont {Glenn}}, \bibinfo
  {author} {\bibfnamefont {S.~J.}\ \bibnamefont {DeVience}}, \bibinfo {author}
  {\bibfnamefont {L.~M.}\ \bibnamefont {Pham}}, \bibinfo {author}
  {\bibfnamefont {L.}~\bibnamefont {Rahn-Lee}}, \bibinfo {author}
  {\bibfnamefont {M.~D.}\ \bibnamefont {Lukin}}, \bibinfo {author}
  {\bibfnamefont {A.}~\bibnamefont {Yacoby}}, \bibinfo {author} {\bibfnamefont
  {A.}~\bibnamefont {Komeili}},\ and\ \bibinfo {author} {\bibfnamefont {R.~L.}\
  \bibnamefont {Walsworth}},\ }\bibfield  {title} {\bibinfo {title} {Optical
  magnetic imaging of living cells},\ }\href
  {https://doi.org/10.1038/nature12072} {\bibfield  {journal} {\bibinfo
  {journal} {Nature}\ }\textbf {\bibinfo {volume} {496}},\ \bibinfo {pages}
  {486} (\bibinfo {year} {2013})}\BibitemShut {NoStop}%
\bibitem [{\citenamefont {Hall}\ \emph {et~al.}(2012)\citenamefont {Hall},
  \citenamefont {Beart}, \citenamefont {Thomas}, \citenamefont {Simpson},
  \citenamefont {McGuinness}, \citenamefont {Cole}, \citenamefont {Manton},
  \citenamefont {Scholten}, \citenamefont {Jelezko}, \citenamefont {Wrachtrup}
  \emph {et~al.}}]{Hall2012}%
  \BibitemOpen
  \bibfield  {author} {\bibinfo {author} {\bibfnamefont {L.}~\bibnamefont
  {Hall}}, \bibinfo {author} {\bibfnamefont {G.}~\bibnamefont {Beart}},
  \bibinfo {author} {\bibfnamefont {E.}~\bibnamefont {Thomas}}, \bibinfo
  {author} {\bibfnamefont {D.}~\bibnamefont {Simpson}}, \bibinfo {author}
  {\bibfnamefont {L.}~\bibnamefont {McGuinness}}, \bibinfo {author}
  {\bibfnamefont {J.}~\bibnamefont {Cole}}, \bibinfo {author} {\bibfnamefont
  {J.}~\bibnamefont {Manton}}, \bibinfo {author} {\bibfnamefont
  {R.}~\bibnamefont {Scholten}}, \bibinfo {author} {\bibfnamefont
  {F.}~\bibnamefont {Jelezko}}, \bibinfo {author} {\bibfnamefont
  {J.}~\bibnamefont {Wrachtrup}}, \emph {et~al.},\ }\bibfield  {title}
  {\bibinfo {title} {High spatial and temporal resolution wide-field imaging of
  neuron activity using quantum nv-diamond},\ }\href
  {https://doi.org/10.1038/srep00401} {\bibfield  {journal} {\bibinfo
  {journal} {Scientific reports}\ }\textbf {\bibinfo {volume} {2}},\ \bibinfo
  {pages} {401} (\bibinfo {year} {2012})}\BibitemShut {NoStop}%
\bibitem [{\citenamefont {Koehl}\ \emph {et~al.}(2011)\citenamefont {Koehl},
  \citenamefont {Buckley}, \citenamefont {Heremans}, \citenamefont {Calusine},\
  and\ \citenamefont {Awschalom}}]{Koehl2011}%
  \BibitemOpen
  \bibfield  {author} {\bibinfo {author} {\bibfnamefont {W.~F.}\ \bibnamefont
  {Koehl}}, \bibinfo {author} {\bibfnamefont {B.~B.}\ \bibnamefont {Buckley}},
  \bibinfo {author} {\bibfnamefont {F.~J.}\ \bibnamefont {Heremans}}, \bibinfo
  {author} {\bibfnamefont {G.}~\bibnamefont {Calusine}},\ and\ \bibinfo
  {author} {\bibfnamefont {D.~D.}\ \bibnamefont {Awschalom}},\ }\bibfield
  {title} {\bibinfo {title} {Room temperature coherent control of defect spin
  qubits in silicon carbide},\ }\href {https://doi.org/10.1038/nature10562}
  {\bibfield  {journal} {\bibinfo  {journal} {Nature}\ }\textbf {\bibinfo
  {volume} {479}},\ \bibinfo {pages} {84} (\bibinfo {year} {2011})}\BibitemShut
  {NoStop}%
\bibitem [{\citenamefont {Awschalom}\ \emph {et~al.}(2018)\citenamefont
  {Awschalom}, \citenamefont {Hanson}, \citenamefont {Wrachtrup},\ and\
  \citenamefont {Zhou}}]{awschalom2018quantum}%
  \BibitemOpen
  \bibfield  {author} {\bibinfo {author} {\bibfnamefont {D.~D.}\ \bibnamefont
  {Awschalom}}, \bibinfo {author} {\bibfnamefont {R.}~\bibnamefont {Hanson}},
  \bibinfo {author} {\bibfnamefont {J.}~\bibnamefont {Wrachtrup}},\ and\
  \bibinfo {author} {\bibfnamefont {B.~B.}\ \bibnamefont {Zhou}},\ }\bibfield
  {title} {\bibinfo {title} {Quantum technologies with optically interfaced
  solid-state spins},\ }\href
  {https://doi.org/https://doi.org/10.1038/s41566-018-0232-2} {\bibfield
  {journal} {\bibinfo  {journal} {Nature Photonics}\ }\textbf {\bibinfo
  {volume} {12}},\ \bibinfo {pages} {516} (\bibinfo {year} {2018})}\BibitemShut
  {NoStop}%
\bibitem [{\citenamefont {Gottscholl}\ \emph {et~al.}(2020)\citenamefont
  {Gottscholl}, \citenamefont {Kianinia}, \citenamefont {Soltamov},
  \citenamefont {Orlinskii}, \citenamefont {Mamin}, \citenamefont {Bradac},
  \citenamefont {Kasper}, \citenamefont {Krambrock}, \citenamefont {Sperlich},
  \citenamefont {Toth}, \citenamefont {Aharonovich},\ and\ \citenamefont
  {Dyakonov}}]{Gottscholl2020}%
  \BibitemOpen
  \bibfield  {author} {\bibinfo {author} {\bibfnamefont {A.}~\bibnamefont
  {Gottscholl}}, \bibinfo {author} {\bibfnamefont {M.}~\bibnamefont
  {Kianinia}}, \bibinfo {author} {\bibfnamefont {V.}~\bibnamefont {Soltamov}},
  \bibinfo {author} {\bibfnamefont {S.}~\bibnamefont {Orlinskii}}, \bibinfo
  {author} {\bibfnamefont {G.}~\bibnamefont {Mamin}}, \bibinfo {author}
  {\bibfnamefont {C.}~\bibnamefont {Bradac}}, \bibinfo {author} {\bibfnamefont
  {C.}~\bibnamefont {Kasper}}, \bibinfo {author} {\bibfnamefont
  {K.}~\bibnamefont {Krambrock}}, \bibinfo {author} {\bibfnamefont
  {A.}~\bibnamefont {Sperlich}}, \bibinfo {author} {\bibfnamefont
  {M.}~\bibnamefont {Toth}}, \bibinfo {author} {\bibfnamefont {I.}~\bibnamefont
  {Aharonovich}},\ and\ \bibinfo {author} {\bibfnamefont {V.}~\bibnamefont
  {Dyakonov}},\ }\bibfield  {title} {\bibinfo {title} {Initialization and
  read-out of intrinsic spin defects in a van der waals crystal at room
  temperature},\ }\href {https://doi.org/10.1038/s41563-020-0619-6} {\bibfield
  {journal} {\bibinfo  {journal} {Nature Materials}\ }\textbf {\bibinfo
  {volume} {19}},\ \bibinfo {pages} {540} (\bibinfo {year} {2020})}\BibitemShut
  {NoStop}%
\bibitem [{\citenamefont {Stern}\ \emph {et~al.}(2024)\citenamefont {Stern},
  \citenamefont {M.~Gilardoni}, \citenamefont {Gu}, \citenamefont
  {Eizagirre~Barker}, \citenamefont {Powell}, \citenamefont {Deng},
  \citenamefont {Fraser}, \citenamefont {Follet}, \citenamefont {Li},
  \citenamefont {Ramsay} \emph {et~al.}}]{stern2024quantum}%
  \BibitemOpen
  \bibfield  {author} {\bibinfo {author} {\bibfnamefont {H.~L.}\ \bibnamefont
  {Stern}}, \bibinfo {author} {\bibfnamefont {C.}~\bibnamefont {M.~Gilardoni}},
  \bibinfo {author} {\bibfnamefont {Q.}~\bibnamefont {Gu}}, \bibinfo {author}
  {\bibfnamefont {S.}~\bibnamefont {Eizagirre~Barker}}, \bibinfo {author}
  {\bibfnamefont {O.~F.}\ \bibnamefont {Powell}}, \bibinfo {author}
  {\bibfnamefont {X.}~\bibnamefont {Deng}}, \bibinfo {author} {\bibfnamefont
  {S.~A.}\ \bibnamefont {Fraser}}, \bibinfo {author} {\bibfnamefont
  {L.}~\bibnamefont {Follet}}, \bibinfo {author} {\bibfnamefont
  {C.}~\bibnamefont {Li}}, \bibinfo {author} {\bibfnamefont {A.~J.}\
  \bibnamefont {Ramsay}}, \emph {et~al.},\ }\bibfield  {title} {\bibinfo
  {title} {A quantum coherent spin in hexagonal boron nitride at ambient
  conditions},\ }\href {https://doi.org/10.1038/s41563-024-01887-z} {\bibfield
  {journal} {\bibinfo  {journal} {Nature Materials}\ }\textbf {\bibinfo
  {volume} {23}},\ \bibinfo {pages} {1379} (\bibinfo {year}
  {2024})}\BibitemShut {NoStop}%
\bibitem [{\citenamefont {Wasielewski}\ \emph {et~al.}(2020)\citenamefont
  {Wasielewski}, \citenamefont {Forbes}, \citenamefont {Frank}, \citenamefont
  {Kowalski}, \citenamefont {Scholes}, \citenamefont {Yuen-Zhou}, \citenamefont
  {Baldo}, \citenamefont {Freedman}, \citenamefont {Goldsmith}, \citenamefont
  {Goodson~III} \emph {et~al.}}]{wasielewski2020exploiting}%
  \BibitemOpen
  \bibfield  {author} {\bibinfo {author} {\bibfnamefont {M.~R.}\ \bibnamefont
  {Wasielewski}}, \bibinfo {author} {\bibfnamefont {M.~D.}\ \bibnamefont
  {Forbes}}, \bibinfo {author} {\bibfnamefont {N.~L.}\ \bibnamefont {Frank}},
  \bibinfo {author} {\bibfnamefont {K.}~\bibnamefont {Kowalski}}, \bibinfo
  {author} {\bibfnamefont {G.~D.}\ \bibnamefont {Scholes}}, \bibinfo {author}
  {\bibfnamefont {J.}~\bibnamefont {Yuen-Zhou}}, \bibinfo {author}
  {\bibfnamefont {M.~A.}\ \bibnamefont {Baldo}}, \bibinfo {author}
  {\bibfnamefont {D.~E.}\ \bibnamefont {Freedman}}, \bibinfo {author}
  {\bibfnamefont {R.~H.}\ \bibnamefont {Goldsmith}}, \bibinfo {author}
  {\bibfnamefont {T.}~\bibnamefont {Goodson~III}}, \emph {et~al.},\ }\bibfield
  {title} {\bibinfo {title} {Exploiting chemistry and molecular systems for
  quantum information science},\ }\href
  {https://doi.org/https://doi.org/10.1038/s41570-020-0200-5} {\bibfield
  {journal} {\bibinfo  {journal} {Nature Reviews Chemistry}\ }\textbf {\bibinfo
  {volume} {4}},\ \bibinfo {pages} {490} (\bibinfo {year} {2020})}\BibitemShut
  {NoStop}%
\bibitem [{\citenamefont {Yu}\ \emph {et~al.}(2021)\citenamefont {Yu},
  \citenamefont {Von~Kugelgen}, \citenamefont {Laorenza},\ and\ \citenamefont
  {Freedman}}]{yu2021molecular}%
  \BibitemOpen
  \bibfield  {author} {\bibinfo {author} {\bibfnamefont {C.-J.}\ \bibnamefont
  {Yu}}, \bibinfo {author} {\bibfnamefont {S.}~\bibnamefont {Von~Kugelgen}},
  \bibinfo {author} {\bibfnamefont {D.~W.}\ \bibnamefont {Laorenza}},\ and\
  \bibinfo {author} {\bibfnamefont {D.~E.}\ \bibnamefont {Freedman}},\
  }\bibfield  {title} {\bibinfo {title} {A molecular approach to quantum
  sensing},\ }\href
  {https://doi.org/https://doi.org/10.1021/acscentsci.0c00737} {\bibfield
  {journal} {\bibinfo  {journal} {ACS central science}\ }\textbf {\bibinfo
  {volume} {7}},\ \bibinfo {pages} {712} (\bibinfo {year} {2021})}\BibitemShut
  {NoStop}%
\bibitem [{\citenamefont {Atzori}\ and\ \citenamefont
  {Sessoli}(2019)}]{atzori2019second}%
  \BibitemOpen
  \bibfield  {author} {\bibinfo {author} {\bibfnamefont {M.}~\bibnamefont
  {Atzori}}\ and\ \bibinfo {author} {\bibfnamefont {R.}~\bibnamefont
  {Sessoli}},\ }\bibfield  {title} {\bibinfo {title} {The second quantum
  revolution: role and challenges of molecular chemistry},\ }\href
  {https://doi.org/10.1021/jacs.9b00984} {\bibfield  {journal} {\bibinfo
  {journal} {Journal of the American Chemical Society}\ }\textbf {\bibinfo
  {volume} {141}},\ \bibinfo {pages} {11339} (\bibinfo {year}
  {2019})}\BibitemShut {NoStop}%
\bibitem [{\citenamefont {Gaita-Ari{\~n}o}\ \emph {et~al.}(2019)\citenamefont
  {Gaita-Ari{\~n}o}, \citenamefont {Luis}, \citenamefont {Hill},\ and\
  \citenamefont {Coronado}}]{Gaita2019}%
  \BibitemOpen
  \bibfield  {author} {\bibinfo {author} {\bibfnamefont {A.}~\bibnamefont
  {Gaita-Ari{\~n}o}}, \bibinfo {author} {\bibfnamefont {F.}~\bibnamefont
  {Luis}}, \bibinfo {author} {\bibfnamefont {S.}~\bibnamefont {Hill}},\ and\
  \bibinfo {author} {\bibfnamefont {E.}~\bibnamefont {Coronado}},\ }\bibfield
  {title} {\bibinfo {title} {Molecular spins for quantum computation},\ }\href
  {https://doi.org/10.1038/s41557-019-0232-y} {\bibfield  {journal} {\bibinfo
  {journal} {Nature chemistry}\ }\textbf {\bibinfo {volume} {11}},\ \bibinfo
  {pages} {301} (\bibinfo {year} {2019})}\BibitemShut {NoStop}%
\bibitem [{\citenamefont {Barry}\ \emph {et~al.}(2020)\citenamefont {Barry},
  \citenamefont {Schloss}, \citenamefont {Bauch}, \citenamefont {Turner},
  \citenamefont {Hart}, \citenamefont {Pham},\ and\ \citenamefont
  {Walsworth}}]{Barry2020}%
  \BibitemOpen
  \bibfield  {author} {\bibinfo {author} {\bibfnamefont {J.~F.}\ \bibnamefont
  {Barry}}, \bibinfo {author} {\bibfnamefont {J.~M.}\ \bibnamefont {Schloss}},
  \bibinfo {author} {\bibfnamefont {E.}~\bibnamefont {Bauch}}, \bibinfo
  {author} {\bibfnamefont {M.~J.}\ \bibnamefont {Turner}}, \bibinfo {author}
  {\bibfnamefont {C.~A.}\ \bibnamefont {Hart}}, \bibinfo {author}
  {\bibfnamefont {L.~M.}\ \bibnamefont {Pham}},\ and\ \bibinfo {author}
  {\bibfnamefont {R.~L.}\ \bibnamefont {Walsworth}},\ }\bibfield  {title}
  {\bibinfo {title} {Sensitivity optimization for nv-diamond magnetometry},\
  }\href {https://doi.org/10.1103/RevModPhys.92.015004} {\bibfield  {journal}
  {\bibinfo  {journal} {Rev. Mod. Phys.}\ }\textbf {\bibinfo {volume} {92}},\
  \bibinfo {pages} {015004} (\bibinfo {year} {2020})}\BibitemShut {NoStop}%
\bibitem [{\citenamefont {Bayliss}\ \emph {et~al.}(2020)\citenamefont
  {Bayliss}, \citenamefont {Laorenza}, \citenamefont {Mintun}, \citenamefont
  {Kovos}, \citenamefont {Freedman},\ and\ \citenamefont
  {Awschalom}}]{bayliss2020optically}%
  \BibitemOpen
  \bibfield  {author} {\bibinfo {author} {\bibfnamefont {S.}~\bibnamefont
  {Bayliss}}, \bibinfo {author} {\bibfnamefont {D.}~\bibnamefont {Laorenza}},
  \bibinfo {author} {\bibfnamefont {P.}~\bibnamefont {Mintun}}, \bibinfo
  {author} {\bibfnamefont {B.}~\bibnamefont {Kovos}}, \bibinfo {author}
  {\bibfnamefont {D.~E.}\ \bibnamefont {Freedman}},\ and\ \bibinfo {author}
  {\bibfnamefont {D.}~\bibnamefont {Awschalom}},\ }\bibfield  {title} {\bibinfo
  {title} {Optically addressable molecular spins for quantum information
  processing},\ }\href {https://doi.org/DOI: 10.1126/science.abb9352}
  {\bibfield  {journal} {\bibinfo  {journal} {Science}\ }\textbf {\bibinfo
  {volume} {370}},\ \bibinfo {pages} {1309} (\bibinfo {year}
  {2020})}\BibitemShut {NoStop}%
\bibitem [{\citenamefont {Serrano}\ \emph {et~al.}(2022)\citenamefont
  {Serrano}, \citenamefont {Kuppusamy}, \citenamefont {Heinrich}, \citenamefont
  {Fuhr}, \citenamefont {Hunger}, \citenamefont {Ruben},\ and\ \citenamefont
  {Goldner}}]{serrano2022ultra}%
  \BibitemOpen
  \bibfield  {author} {\bibinfo {author} {\bibfnamefont {D.}~\bibnamefont
  {Serrano}}, \bibinfo {author} {\bibfnamefont {S.~K.}\ \bibnamefont
  {Kuppusamy}}, \bibinfo {author} {\bibfnamefont {B.}~\bibnamefont {Heinrich}},
  \bibinfo {author} {\bibfnamefont {O.}~\bibnamefont {Fuhr}}, \bibinfo {author}
  {\bibfnamefont {D.}~\bibnamefont {Hunger}}, \bibinfo {author} {\bibfnamefont
  {M.}~\bibnamefont {Ruben}},\ and\ \bibinfo {author} {\bibfnamefont
  {P.}~\bibnamefont {Goldner}},\ }\bibfield  {title} {\bibinfo {title}
  {Ultra-narrow optical linewidths in rare-earth molecular crystals},\ }\href
  {https://doi.org/https://doi.org/10.1038/s41586-021-04316-2} {\bibfield
  {journal} {\bibinfo  {journal} {Nature}\ }\textbf {\bibinfo {volume} {603}},\
  \bibinfo {pages} {241} (\bibinfo {year} {2022})}\BibitemShut {NoStop}%
\bibitem [{\citenamefont {Gorgon}\ \emph {et~al.}(2023)\citenamefont {Gorgon},
  \citenamefont {Lv}, \citenamefont {Gr{\"u}ne}, \citenamefont {Drummond},
  \citenamefont {Myers}, \citenamefont {Londi}, \citenamefont {Ricci},
  \citenamefont {Valverde}, \citenamefont {Tonnel{\'e}}, \citenamefont {Murto}
  \emph {et~al.}}]{gorgon2023reversible}%
  \BibitemOpen
  \bibfield  {author} {\bibinfo {author} {\bibfnamefont {S.}~\bibnamefont
  {Gorgon}}, \bibinfo {author} {\bibfnamefont {K.}~\bibnamefont {Lv}}, \bibinfo
  {author} {\bibfnamefont {J.}~\bibnamefont {Gr{\"u}ne}}, \bibinfo {author}
  {\bibfnamefont {B.~H.}\ \bibnamefont {Drummond}}, \bibinfo {author}
  {\bibfnamefont {W.~K.}\ \bibnamefont {Myers}}, \bibinfo {author}
  {\bibfnamefont {G.}~\bibnamefont {Londi}}, \bibinfo {author} {\bibfnamefont
  {G.}~\bibnamefont {Ricci}}, \bibinfo {author} {\bibfnamefont
  {D.}~\bibnamefont {Valverde}}, \bibinfo {author} {\bibfnamefont
  {C.}~\bibnamefont {Tonnel{\'e}}}, \bibinfo {author} {\bibfnamefont
  {P.}~\bibnamefont {Murto}}, \emph {et~al.},\ }\bibfield  {title} {\bibinfo
  {title} {Reversible spin-optical interface in luminescent organic radicals},\
  }\href {https://doi.org/https://doi.org/10.1038/s41586-023-06222-1}
  {\bibfield  {journal} {\bibinfo  {journal} {Nature}\ }\textbf {\bibinfo
  {volume} {620}},\ \bibinfo {pages} {538} (\bibinfo {year}
  {2023})}\BibitemShut {NoStop}%
\bibitem [{\citenamefont {Quintes}\ \emph {et~al.}(2023)\citenamefont
  {Quintes}, \citenamefont {Mayl{\"a}nder},\ and\ \citenamefont
  {Richert}}]{quintes2023properties}%
  \BibitemOpen
  \bibfield  {author} {\bibinfo {author} {\bibfnamefont {T.}~\bibnamefont
  {Quintes}}, \bibinfo {author} {\bibfnamefont {M.}~\bibnamefont
  {Mayl{\"a}nder}},\ and\ \bibinfo {author} {\bibfnamefont {S.}~\bibnamefont
  {Richert}},\ }\bibfield  {title} {\bibinfo {title} {Properties and
  applications of photoexcited chromophore--radical systems},\ }\href
  {https://doi.org/https://doi.org/10.1038/s41570-022-00453-y} {\bibfield
  {journal} {\bibinfo  {journal} {Nature Reviews Chemistry}\ }\textbf {\bibinfo
  {volume} {7}},\ \bibinfo {pages} {75} (\bibinfo {year} {2023})}\BibitemShut
  {NoStop}%
\bibitem [{\citenamefont {Kopp}\ \emph {et~al.}(2024)\citenamefont {Kopp},
  \citenamefont {Nakamura}, \citenamefont {Phelan}, \citenamefont {Poh},
  \citenamefont {Tyndall}, \citenamefont {Brown}, \citenamefont {Huang},
  \citenamefont {Yuen-Zhou}, \citenamefont {Krzyaniak},\ and\ \citenamefont
  {Wasielewski}}]{kopp2024luminescent}%
  \BibitemOpen
  \bibfield  {author} {\bibinfo {author} {\bibfnamefont {S.~M.}\ \bibnamefont
  {Kopp}}, \bibinfo {author} {\bibfnamefont {S.}~\bibnamefont {Nakamura}},
  \bibinfo {author} {\bibfnamefont {B.~T.}\ \bibnamefont {Phelan}}, \bibinfo
  {author} {\bibfnamefont {Y.~R.}\ \bibnamefont {Poh}}, \bibinfo {author}
  {\bibfnamefont {S.~B.}\ \bibnamefont {Tyndall}}, \bibinfo {author}
  {\bibfnamefont {P.~J.}\ \bibnamefont {Brown}}, \bibinfo {author}
  {\bibfnamefont {Y.}~\bibnamefont {Huang}}, \bibinfo {author} {\bibfnamefont
  {J.}~\bibnamefont {Yuen-Zhou}}, \bibinfo {author} {\bibfnamefont {M.~D.}\
  \bibnamefont {Krzyaniak}},\ and\ \bibinfo {author} {\bibfnamefont {M.~R.}\
  \bibnamefont {Wasielewski}},\ }\bibfield  {title} {\bibinfo {title}
  {Luminescent organic triplet diradicals as optically addressable molecular
  qubits},\ }\href {https://doi.org/DOI: 10.1021/jacs.4c11116} {\bibfield
  {journal} {\bibinfo  {journal} {Journal of the American Chemical Society}\
  }\textbf {\bibinfo {volume} {146}},\ \bibinfo {pages} {27935} (\bibinfo
  {year} {2024})}\BibitemShut {NoStop}%
\bibitem [{\citenamefont {Chowdhury}\ \emph {et~al.}(2024)\citenamefont
  {Chowdhury}, \citenamefont {Murto}, \citenamefont {Panjwani}, \citenamefont
  {Sun}, \citenamefont {Ghosh}, \citenamefont {Boeije}, \citenamefont
  {Derkach}, \citenamefont {Woo}, \citenamefont {Millington}, \citenamefont
  {Congrave} \emph {et~al.}}]{chowdhury2024optical}%
  \BibitemOpen
  \bibfield  {author} {\bibinfo {author} {\bibfnamefont {R.}~\bibnamefont
  {Chowdhury}}, \bibinfo {author} {\bibfnamefont {P.}~\bibnamefont {Murto}},
  \bibinfo {author} {\bibfnamefont {N.~A.}\ \bibnamefont {Panjwani}}, \bibinfo
  {author} {\bibfnamefont {Y.}~\bibnamefont {Sun}}, \bibinfo {author}
  {\bibfnamefont {P.}~\bibnamefont {Ghosh}}, \bibinfo {author} {\bibfnamefont
  {Y.}~\bibnamefont {Boeije}}, \bibinfo {author} {\bibfnamefont
  {V.}~\bibnamefont {Derkach}}, \bibinfo {author} {\bibfnamefont {S.-J.}\
  \bibnamefont {Woo}}, \bibinfo {author} {\bibfnamefont {O.}~\bibnamefont
  {Millington}}, \bibinfo {author} {\bibfnamefont {D.~G.}\ \bibnamefont
  {Congrave}}, \emph {et~al.},\ }\bibfield  {title} {\bibinfo {title} {Optical
  read and write of spin states in organic diradicals},\ }\bibfield  {journal}
  {\bibinfo  {journal} {arXiv preprint arXiv:2406.03365}\ }\href
  {https://doi.org/https://doi.org/10.48550/arXiv.2406.03365}
  {https://doi.org/10.48550/arXiv.2406.03365} (\bibinfo {year}
  {2024})\BibitemShut {NoStop}%
\bibitem [{\citenamefont {Sutcliffe}\ \emph {et~al.}(2024)\citenamefont
  {Sutcliffe}, \citenamefont {Kazmierczak},\ and\ \citenamefont
  {Hadt}}]{sutcliffe2024ultrafast}%
  \BibitemOpen
  \bibfield  {author} {\bibinfo {author} {\bibfnamefont {E.}~\bibnamefont
  {Sutcliffe}}, \bibinfo {author} {\bibfnamefont {N.~P.}\ \bibnamefont
  {Kazmierczak}},\ and\ \bibinfo {author} {\bibfnamefont {R.~G.}\ \bibnamefont
  {Hadt}},\ }\bibfield  {title} {\bibinfo {title} {Ultrafast all-optical
  coherence of molecular electron spins in room-temperature water solution},\
  }\href {https://doi.org/DOI: 10.1126/science.ads051} {\bibfield  {journal}
  {\bibinfo  {journal} {Science}\ }\textbf {\bibinfo {volume} {386}},\ \bibinfo
  {pages} {888} (\bibinfo {year} {2024})}\BibitemShut {NoStop}%
\bibitem [{\citenamefont {Shin}\ \emph {et~al.}(2024)\citenamefont {Shin},
  \citenamefont {Zhao}, \citenamefont {Shen}, \citenamefont {Dickerson},
  \citenamefont {Li}, \citenamefont {Roshandel}, \citenamefont {B{\'\i}m},
  \citenamefont {Atallah}, \citenamefont {Oyala}, \citenamefont {He} \emph
  {et~al.}}]{shin2024toward}%
  \BibitemOpen
  \bibfield  {author} {\bibinfo {author} {\bibfnamefont {A.~J.}\ \bibnamefont
  {Shin}}, \bibinfo {author} {\bibfnamefont {C.}~\bibnamefont {Zhao}}, \bibinfo
  {author} {\bibfnamefont {Y.}~\bibnamefont {Shen}}, \bibinfo {author}
  {\bibfnamefont {C.~E.}\ \bibnamefont {Dickerson}}, \bibinfo {author}
  {\bibfnamefont {B.}~\bibnamefont {Li}}, \bibinfo {author} {\bibfnamefont
  {H.}~\bibnamefont {Roshandel}}, \bibinfo {author} {\bibfnamefont
  {D.}~\bibnamefont {B{\'\i}m}}, \bibinfo {author} {\bibfnamefont {T.~L.}\
  \bibnamefont {Atallah}}, \bibinfo {author} {\bibfnamefont {P.~H.}\
  \bibnamefont {Oyala}}, \bibinfo {author} {\bibfnamefont {Y.}~\bibnamefont
  {He}}, \emph {et~al.},\ }\bibfield  {title} {\bibinfo {title} {Toward liquid
  cell quantum sensing: Ytterbium complexes with ultranarrow absorption},\
  }\href {https://doi.org/10.1126/science.adf7577} {\bibfield  {journal}
  {\bibinfo  {journal} {Science}\ }\textbf {\bibinfo {volume} {385}},\ \bibinfo
  {pages} {651} (\bibinfo {year} {2024})}\BibitemShut {NoStop}%
\bibitem [{\citenamefont {Sloop}\ \emph {et~al.}(1981)\citenamefont {Sloop},
  \citenamefont {Yu}, \citenamefont {Lin},\ and\ \citenamefont
  {Weissman}}]{Sloop1981}%
  \BibitemOpen
  \bibfield  {author} {\bibinfo {author} {\bibfnamefont {D.~J.}\ \bibnamefont
  {Sloop}}, \bibinfo {author} {\bibfnamefont {H.-L.}\ \bibnamefont {Yu}},
  \bibinfo {author} {\bibfnamefont {T.-S.}\ \bibnamefont {Lin}},\ and\ \bibinfo
  {author} {\bibfnamefont {S.~I.}\ \bibnamefont {Weissman}},\ }\bibfield
  {title} {\bibinfo {title} {Electron spin echoes of a photoexcited triplet:
  Pentacene in \textit{p}-terphenyl crystals},\ }\href
  {https://doi.org/10.1063/1.442520} {\bibfield  {journal} {\bibinfo  {journal}
  {The Journal of Chemical Physics}\ }\textbf {\bibinfo {volume} {75}},\
  \bibinfo {pages} {3746} (\bibinfo {year} {1981})}\BibitemShut {NoStop}%
\bibitem [{\citenamefont {Wrachtrup}\ \emph {et~al.}(1993)\citenamefont
  {Wrachtrup}, \citenamefont {von Borczyskowski}, \citenamefont {Bernard},
  \citenamefont {Orrit},\ and\ \citenamefont {Brown}}]{Wrachtrup1993}%
  \BibitemOpen
  \bibfield  {author} {\bibinfo {author} {\bibfnamefont {J.}~\bibnamefont
  {Wrachtrup}}, \bibinfo {author} {\bibfnamefont {C.}~\bibnamefont {von
  Borczyskowski}}, \bibinfo {author} {\bibfnamefont {J.}~\bibnamefont
  {Bernard}}, \bibinfo {author} {\bibfnamefont {M.}~\bibnamefont {Orrit}},\
  and\ \bibinfo {author} {\bibfnamefont {R.}~\bibnamefont {Brown}},\ }\bibfield
   {title} {\bibinfo {title} {Optical detection of magnetic resonance in a
  single molecule},\ }\href {https://doi.org/10.1038/363244a0} {\bibfield
  {journal} {\bibinfo  {journal} {Nature}\ }\textbf {\bibinfo {volume} {363}},\
  \bibinfo {pages} {244} (\bibinfo {year} {1993})}\BibitemShut {NoStop}%
\bibitem [{\citenamefont {Köhler}\ \emph {et~al.}(1993)\citenamefont
  {Köhler}, \citenamefont {Disselhorst}, \citenamefont {Donckers},
  \citenamefont {Groenen}, \citenamefont {Schmidt},\ and\ \citenamefont
  {Moerner}}]{Kohler1993}%
  \BibitemOpen
  \bibfield  {author} {\bibinfo {author} {\bibfnamefont {J.}~\bibnamefont
  {Köhler}}, \bibinfo {author} {\bibfnamefont {J.~A. J.~M.}\ \bibnamefont
  {Disselhorst}}, \bibinfo {author} {\bibfnamefont {M.~C. J.~M.}\ \bibnamefont
  {Donckers}}, \bibinfo {author} {\bibfnamefont {E.~J.~J.}\ \bibnamefont
  {Groenen}}, \bibinfo {author} {\bibfnamefont {J.}~\bibnamefont {Schmidt}},\
  and\ \bibinfo {author} {\bibfnamefont {W.~E.}\ \bibnamefont {Moerner}},\
  }\bibfield  {title} {\bibinfo {title} {Magnetic resonance of a single
  molecular spin},\ }\href {https://doi.org/10.1038/363242a0} {\bibfield
  {journal} {\bibinfo  {journal} {Nature}\ }\textbf {\bibinfo {volume} {363}},\
  \bibinfo {pages} {242} (\bibinfo {year} {1993})}\BibitemShut {NoStop}%
\bibitem [{\citenamefont {Moro}\ \emph {et~al.}(2022)\citenamefont {Moro},
  \citenamefont {Moret}, \citenamefont {Ghirri}, \citenamefont {Granados~del
  {\'A}guila}, \citenamefont {Kubozono}, \citenamefont {Beverina},\ and\
  \citenamefont {Cassinese}}]{moro2022room}%
  \BibitemOpen
  \bibfield  {author} {\bibinfo {author} {\bibfnamefont {F.}~\bibnamefont
  {Moro}}, \bibinfo {author} {\bibfnamefont {M.}~\bibnamefont {Moret}},
  \bibinfo {author} {\bibfnamefont {A.}~\bibnamefont {Ghirri}}, \bibinfo
  {author} {\bibfnamefont {A.}~\bibnamefont {Granados~del {\'A}guila}},
  \bibinfo {author} {\bibfnamefont {Y.}~\bibnamefont {Kubozono}}, \bibinfo
  {author} {\bibfnamefont {L.}~\bibnamefont {Beverina}},\ and\ \bibinfo
  {author} {\bibfnamefont {A.}~\bibnamefont {Cassinese}},\ }\bibfield  {title}
  {\bibinfo {title} {Room-temperature optically detected magnetic resonance of
  triplet excitons in a pentacene-doped picene single crystal},\ }\href
  {https://doi.org/10.1557/s43578-022-00536-y} {\bibfield  {journal} {\bibinfo
  {journal} {Journal of Materials Research}\ }\textbf {\bibinfo {volume}
  {37}},\ \bibinfo {pages} {1269} (\bibinfo {year} {2022})}\BibitemShut
  {NoStop}%
\bibitem [{\citenamefont {Mena}\ \emph {et~al.}(2024)\citenamefont {Mena},
  \citenamefont {Mann}, \citenamefont {Cowley-Semple}, \citenamefont {Bryan},
  \citenamefont {Heutz}, \citenamefont {McCamey}, \citenamefont {Attwood},\
  and\ \citenamefont {Bayliss}}]{Mena2024}%
  \BibitemOpen
  \bibfield  {author} {\bibinfo {author} {\bibfnamefont {A.}~\bibnamefont
  {Mena}}, \bibinfo {author} {\bibfnamefont {S.~K.}\ \bibnamefont {Mann}},
  \bibinfo {author} {\bibfnamefont {A.}~\bibnamefont {Cowley-Semple}}, \bibinfo
  {author} {\bibfnamefont {E.}~\bibnamefont {Bryan}}, \bibinfo {author}
  {\bibfnamefont {S.}~\bibnamefont {Heutz}}, \bibinfo {author} {\bibfnamefont
  {D.~R.}\ \bibnamefont {McCamey}}, \bibinfo {author} {\bibfnamefont
  {M.}~\bibnamefont {Attwood}},\ and\ \bibinfo {author} {\bibfnamefont {S.~L.}\
  \bibnamefont {Bayliss}},\ }\bibfield  {title} {\bibinfo {title}
  {Room-temperature optically detected coherent control of molecular spins},\
  }\href {https://doi.org/10.1103/PhysRevLett.133.120801} {\bibfield  {journal}
  {\bibinfo  {journal} {Phys. Rev. Lett.}\ }\textbf {\bibinfo {volume} {133}},\
  \bibinfo {pages} {120801} (\bibinfo {year} {2024})}\BibitemShut {NoStop}%
\bibitem [{\citenamefont {Singh}\ \emph {et~al.}(2025)\citenamefont {Singh},
  \citenamefont {D'Souza}, \citenamefont {Zhong}, \citenamefont {Druga},
  \citenamefont {Oshiro}, \citenamefont {Blankenship}, \citenamefont {Montis},
  \citenamefont {Reimer}, \citenamefont {Breeze},\ and\ \citenamefont
  {Ajoy}}]{Singh2025}%
  \BibitemOpen
  \bibfield  {author} {\bibinfo {author} {\bibfnamefont {H.}~\bibnamefont
  {Singh}}, \bibinfo {author} {\bibfnamefont {N.}~\bibnamefont {D'Souza}},
  \bibinfo {author} {\bibfnamefont {K.}~\bibnamefont {Zhong}}, \bibinfo
  {author} {\bibfnamefont {E.}~\bibnamefont {Druga}}, \bibinfo {author}
  {\bibfnamefont {J.}~\bibnamefont {Oshiro}}, \bibinfo {author} {\bibfnamefont
  {B.}~\bibnamefont {Blankenship}}, \bibinfo {author} {\bibfnamefont
  {R.}~\bibnamefont {Montis}}, \bibinfo {author} {\bibfnamefont {J.~A.}\
  \bibnamefont {Reimer}}, \bibinfo {author} {\bibfnamefont {J.~D.}\
  \bibnamefont {Breeze}},\ and\ \bibinfo {author} {\bibfnamefont
  {A.}~\bibnamefont {Ajoy}},\ }\bibfield  {title} {\bibinfo {title}
  {Room-temperature quantum sensing with photoexcited triplet electrons in
  organic crystals},\ }\href {https://doi.org/10.1103/PhysRevResearch.7.013192}
  {\bibfield  {journal} {\bibinfo  {journal} {Phys. Rev. Res.}\ }\textbf
  {\bibinfo {volume} {7}},\ \bibinfo {pages} {013192} (\bibinfo {year}
  {2025})}\BibitemShut {NoStop}%
\bibitem [{\citenamefont {Feder}\ \emph {et~al.}(2024)\citenamefont {Feder},
  \citenamefont {Soloway}, \citenamefont {Verma}, \citenamefont {Geng},
  \citenamefont {Wang}, \citenamefont {Kifle}, \citenamefont {Riendeau},
  \citenamefont {Tsaturyan}, \citenamefont {Weiss}, \citenamefont {Xie} \emph
  {et~al.}}]{Feder2024}%
  \BibitemOpen
  \bibfield  {author} {\bibinfo {author} {\bibfnamefont {J.~S.}\ \bibnamefont
  {Feder}}, \bibinfo {author} {\bibfnamefont {B.~S.}\ \bibnamefont {Soloway}},
  \bibinfo {author} {\bibfnamefont {S.}~\bibnamefont {Verma}}, \bibinfo
  {author} {\bibfnamefont {Z.~Z.}\ \bibnamefont {Geng}}, \bibinfo {author}
  {\bibfnamefont {S.}~\bibnamefont {Wang}}, \bibinfo {author} {\bibfnamefont
  {B.}~\bibnamefont {Kifle}}, \bibinfo {author} {\bibfnamefont {E.~G.}\
  \bibnamefont {Riendeau}}, \bibinfo {author} {\bibfnamefont {Y.}~\bibnamefont
  {Tsaturyan}}, \bibinfo {author} {\bibfnamefont {L.~R.}\ \bibnamefont
  {Weiss}}, \bibinfo {author} {\bibfnamefont {M.}~\bibnamefont {Xie}}, \emph
  {et~al.},\ }\bibfield  {title} {\bibinfo {title} {A fluorescent-protein spin
  qubit},\ }\bibfield  {journal} {\bibinfo  {journal} {arXiv preprint
  arXiv:2411.16835}\ }\href
  {https://doi.org/https://doi.org/10.48550/arXiv.2411.16835}
  {https://doi.org/10.48550/arXiv.2411.16835} (\bibinfo {year}
  {2024})\BibitemShut {NoStop}%
\bibitem [{\citenamefont {Abrahams}\ \emph {et~al.}(2024)\citenamefont
  {Abrahams}, \citenamefont {Spreng}, \citenamefont {{\v{S}}tuhec},
  \citenamefont {Kempf}, \citenamefont {James}, \citenamefont {Sechkar},
  \citenamefont {Stacey}, \citenamefont {Trelles-Fernandez}, \citenamefont
  {Antill}, \citenamefont {Ingaramo} \emph {et~al.}}]{abrahams2024quantum}%
  \BibitemOpen
  \bibfield  {author} {\bibinfo {author} {\bibfnamefont {G.}~\bibnamefont
  {Abrahams}}, \bibinfo {author} {\bibfnamefont {V.}~\bibnamefont {Spreng}},
  \bibinfo {author} {\bibfnamefont {A.}~\bibnamefont {{\v{S}}tuhec}}, \bibinfo
  {author} {\bibfnamefont {I.}~\bibnamefont {Kempf}}, \bibinfo {author}
  {\bibfnamefont {J.}~\bibnamefont {James}}, \bibinfo {author} {\bibfnamefont
  {K.}~\bibnamefont {Sechkar}}, \bibinfo {author} {\bibfnamefont
  {S.}~\bibnamefont {Stacey}}, \bibinfo {author} {\bibfnamefont
  {V.}~\bibnamefont {Trelles-Fernandez}}, \bibinfo {author} {\bibfnamefont
  {L.~M.}\ \bibnamefont {Antill}}, \bibinfo {author} {\bibfnamefont
  {M.}~\bibnamefont {Ingaramo}}, \emph {et~al.},\ }\bibfield  {title} {\bibinfo
  {title} {Quantum spin resonance in engineered magneto-sensitive fluorescent
  proteins enables multi-modal sensing in living cells},\ }\bibfield  {journal}
  {\bibinfo  {journal} {bioRxiv}\ }\href
  {https://doi.org/https://doi.org/10.1101/2024.11.25.625143}
  {https://doi.org/10.1101/2024.11.25.625143} (\bibinfo {year}
  {2024})\BibitemShut {NoStop}%
\bibitem [{\citenamefont {Di~Valentin}\ \emph {et~al.}(2014)\citenamefont
  {Di~Valentin}, \citenamefont {Albertini}, \citenamefont {Zurlo},
  \citenamefont {Gobbo},\ and\ \citenamefont {Carbonera}}]{valentin2014}%
  \BibitemOpen
  \bibfield  {author} {\bibinfo {author} {\bibfnamefont {M.}~\bibnamefont
  {Di~Valentin}}, \bibinfo {author} {\bibfnamefont {M.}~\bibnamefont
  {Albertini}}, \bibinfo {author} {\bibfnamefont {E.}~\bibnamefont {Zurlo}},
  \bibinfo {author} {\bibfnamefont {M.}~\bibnamefont {Gobbo}},\ and\ \bibinfo
  {author} {\bibfnamefont {D.}~\bibnamefont {Carbonera}},\ }\bibfield  {title}
  {\bibinfo {title} {Porphyrin triplet state as a potential spin label for
  nanometer distance measurements by {PELDOR} spectroscopy},\ }\href
  {https://doi.org/10.1021/ja502615n} {\bibfield  {journal} {\bibinfo
  {journal} {Journal of the American Chemical Society}\ }\textbf {\bibinfo
  {volume} {136}},\ \bibinfo {pages} {6582} (\bibinfo {year}
  {2014})}\BibitemShut {NoStop}%
\bibitem [{\citenamefont {Hintze}\ \emph {et~al.}(2016)\citenamefont {Hintze},
  \citenamefont {Bu{\"u}cker}, \citenamefont {Domingo~K{\"o}hler},
  \citenamefont {Jeschke},\ and\ \citenamefont {Drescher}}]{hintze2016laser}%
  \BibitemOpen
  \bibfield  {author} {\bibinfo {author} {\bibfnamefont {C.}~\bibnamefont
  {Hintze}}, \bibinfo {author} {\bibfnamefont {D.}~\bibnamefont {Bu{\"u}cker}},
  \bibinfo {author} {\bibfnamefont {S.}~\bibnamefont {Domingo~K{\"o}hler}},
  \bibinfo {author} {\bibfnamefont {G.}~\bibnamefont {Jeschke}},\ and\ \bibinfo
  {author} {\bibfnamefont {M.}~\bibnamefont {Drescher}},\ }\bibfield  {title}
  {\bibinfo {title} {Laser-induced magnetic dipole spectroscopy},\ }\href
  {https://doi.org/DOI: 10.1021/acs.jpclett.6b00765} {\bibfield  {journal}
  {\bibinfo  {journal} {The journal of physical chemistry letters}\ }\textbf
  {\bibinfo {volume} {7}},\ \bibinfo {pages} {2204} (\bibinfo {year}
  {2016})}\BibitemShut {NoStop}%
\bibitem [{\citenamefont {Bertran}\ \emph {et~al.}(2021)\citenamefont
  {Bertran}, \citenamefont {Henbest}, \citenamefont {Zotti}, \citenamefont
  {Gobbo}, \citenamefont {Timmel}, \citenamefont {Valentin},\ and\
  \citenamefont {Bowen}}]{Bertran2021}%
  \BibitemOpen
  \bibfield  {author} {\bibinfo {author} {\bibfnamefont {A.}~\bibnamefont
  {Bertran}}, \bibinfo {author} {\bibfnamefont {K.~B.}\ \bibnamefont
  {Henbest}}, \bibinfo {author} {\bibfnamefont {M.~D.}\ \bibnamefont {Zotti}},
  \bibinfo {author} {\bibfnamefont {M.}~\bibnamefont {Gobbo}}, \bibinfo
  {author} {\bibfnamefont {C.~R.}\ \bibnamefont {Timmel}}, \bibinfo {author}
  {\bibfnamefont {M.~D.}\ \bibnamefont {Valentin}},\ and\ \bibinfo {author}
  {\bibfnamefont {A.~M.}\ \bibnamefont {Bowen}},\ }\bibfield  {title} {\bibinfo
  {title} {Light-induced triplet–triplet electron resonance spectroscopy},\
  }\href {https://doi.org/10.1021/acs.jpclett.0c02884} {\bibfield  {journal}
  {\bibinfo  {journal} {The Journal of Physical Chemistry Letters}\ }\textbf
  {\bibinfo {volume} {12}},\ \bibinfo {pages} {80} (\bibinfo {year}
  {2021})}\BibitemShut {NoStop}%
\bibitem [{\citenamefont {Pansare}\ \emph {et~al.}(2014)\citenamefont
  {Pansare}, \citenamefont {Bruzek}, \citenamefont {Adamson}, \citenamefont
  {Anthony},\ and\ \citenamefont {Prud'homme}}]{Pansare2014}%
  \BibitemOpen
  \bibfield  {author} {\bibinfo {author} {\bibfnamefont {V.~J.}\ \bibnamefont
  {Pansare}}, \bibinfo {author} {\bibfnamefont {M.~J.}\ \bibnamefont {Bruzek}},
  \bibinfo {author} {\bibfnamefont {D.~H.}\ \bibnamefont {Adamson}}, \bibinfo
  {author} {\bibfnamefont {J.}~\bibnamefont {Anthony}},\ and\ \bibinfo {author}
  {\bibfnamefont {R.~K.}\ \bibnamefont {Prud'homme}},\ }\bibfield  {title}
  {\bibinfo {title} {Composite fluorescent nanoparticles for biomedical
  imaging},\ }\href {https://doi.org/10.1007/s11307-013-0689-9} {\bibfield
  {journal} {\bibinfo  {journal} {Molecular Imaging and Biology}\ }\textbf
  {\bibinfo {volume} {16}},\ \bibinfo {pages} {180} (\bibinfo {year}
  {2014})}\BibitemShut {NoStop}%
\bibitem [{\citenamefont {Ishiwata}\ \emph {et~al.}(2025)\citenamefont
  {Ishiwata}, \citenamefont {Song}, \citenamefont {Shigeno}, \citenamefont
  {Nishimura},\ and\ \citenamefont {Yanai}}]{Ishiwata2025}%
  \BibitemOpen
  \bibfield  {author} {\bibinfo {author} {\bibfnamefont {H.}~\bibnamefont
  {Ishiwata}}, \bibinfo {author} {\bibfnamefont {J.}~\bibnamefont {Song}},
  \bibinfo {author} {\bibfnamefont {Y.}~\bibnamefont {Shigeno}}, \bibinfo
  {author} {\bibfnamefont {K.}~\bibnamefont {Nishimura}},\ and\ \bibinfo
  {author} {\bibfnamefont {N.}~\bibnamefont {Yanai}},\ }\bibfield  {title}
  {\bibinfo {title} {Molecular quantum nanosensors functioning in living
  cells},\ }\bibfield  {journal} {\bibinfo  {journal} {ChemRxiv preprint}\
  }\href {https://doi.org/10.26434/chemrxiv-2025-k7db1}
  {10.26434/chemrxiv-2025-k7db1} (\bibinfo {year} {2025})\BibitemShut {NoStop}%
\bibitem [{\citenamefont {Ng}\ \emph {et~al.}(2023)\citenamefont {Ng},
  \citenamefont {Xu}, \citenamefont {Attwood}, \citenamefont {Wu},
  \citenamefont {Meng}, \citenamefont {Chen},\ and\ \citenamefont
  {Oxborrow}}]{Ng2023}%
  \BibitemOpen
  \bibfield  {author} {\bibinfo {author} {\bibfnamefont {W.}~\bibnamefont
  {Ng}}, \bibinfo {author} {\bibfnamefont {X.}~\bibnamefont {Xu}}, \bibinfo
  {author} {\bibfnamefont {M.}~\bibnamefont {Attwood}}, \bibinfo {author}
  {\bibfnamefont {H.}~\bibnamefont {Wu}}, \bibinfo {author} {\bibfnamefont
  {Z.}~\bibnamefont {Meng}}, \bibinfo {author} {\bibfnamefont {X.}~\bibnamefont
  {Chen}},\ and\ \bibinfo {author} {\bibfnamefont {M.}~\bibnamefont
  {Oxborrow}},\ }\bibfield  {title} {\bibinfo {title} {Move aside pentacene:
  Diazapentacene-doped para-terphenyl, a zero-field room-temperature maser with
  strong coupling for cavity quantum electrodynamics},\ }\href
  {https://doi.org/https://doi.org/10.1002/adma.202300441} {\bibfield
  {journal} {\bibinfo  {journal} {Advanced Materials}\ }\textbf {\bibinfo
  {volume} {35}},\ \bibinfo {pages} {2300441} (\bibinfo {year}
  {2023})}\BibitemShut {NoStop}%
\bibitem [{\citenamefont {Kouno}\ \emph {et~al.}(2019)\citenamefont {Kouno},
  \citenamefont {Kawashima}, \citenamefont {Tateishi}, \citenamefont {Uesaka},
  \citenamefont {Kimizuka},\ and\ \citenamefont {Yanai}}]{Kouno2019}%
  \BibitemOpen
  \bibfield  {author} {\bibinfo {author} {\bibfnamefont {H.}~\bibnamefont
  {Kouno}}, \bibinfo {author} {\bibfnamefont {Y.}~\bibnamefont {Kawashima}},
  \bibinfo {author} {\bibfnamefont {K.}~\bibnamefont {Tateishi}}, \bibinfo
  {author} {\bibfnamefont {T.}~\bibnamefont {Uesaka}}, \bibinfo {author}
  {\bibfnamefont {N.}~\bibnamefont {Kimizuka}},\ and\ \bibinfo {author}
  {\bibfnamefont {N.}~\bibnamefont {Yanai}},\ }\bibfield  {title} {\bibinfo
  {title} {Nonpentacene polarizing agents with improved air stability for
  triplet dynamic nuclear polarization at room temperature},\ }\href
  {https://doi.org/10.1021/acs.jpclett.9b00480} {\bibfield  {journal} {\bibinfo
   {journal} {The Journal of Physical Chemistry Letters}\ }\textbf {\bibinfo
  {volume} {10}},\ \bibinfo {pages} {2208} (\bibinfo {year} {2019})},\ \bibinfo
  {note} {pMID: 30933529}\BibitemShut {NoStop}%
\bibitem [{\citenamefont {Takeda}\ \emph {et~al.}(2002)\citenamefont {Takeda},
  \citenamefont {Takegoshi},\ and\ \citenamefont {Terao}}]{Takeda2002}%
  \BibitemOpen
  \bibfield  {author} {\bibinfo {author} {\bibfnamefont {K.}~\bibnamefont
  {Takeda}}, \bibinfo {author} {\bibfnamefont {K.}~\bibnamefont {Takegoshi}},\
  and\ \bibinfo {author} {\bibfnamefont {T.}~\bibnamefont {Terao}},\ }\bibfield
   {title} {\bibinfo {title} {Zero-field electron spin resonance and
  theoretical studies of light penetration into single crystal and
  polycrystalline material doped with molecules photoexcitable to the triplet
  state via intersystem crossing},\ }\href {https://doi.org/10.1063/1.1499124}
  {\bibfield  {journal} {\bibinfo  {journal} {The Journal of Chemical Physics}\
  }\textbf {\bibinfo {volume} {117}},\ \bibinfo {pages} {4940} (\bibinfo {year}
  {2002})}\BibitemShut {NoStop}%
\bibitem [{\citenamefont {Wirtitsch}\ \emph {et~al.}(2023)\citenamefont
  {Wirtitsch}, \citenamefont {Wachter}, \citenamefont {Reisenbauer},
  \citenamefont {Gulka}, \citenamefont {Iv\'ady}, \citenamefont {Jelezko},
  \citenamefont {Gali}, \citenamefont {Nesladek},\ and\ \citenamefont
  {Trupke}}]{Wirtitsch2023}%
  \BibitemOpen
  \bibfield  {author} {\bibinfo {author} {\bibfnamefont {D.}~\bibnamefont
  {Wirtitsch}}, \bibinfo {author} {\bibfnamefont {G.}~\bibnamefont {Wachter}},
  \bibinfo {author} {\bibfnamefont {S.}~\bibnamefont {Reisenbauer}}, \bibinfo
  {author} {\bibfnamefont {M.}~\bibnamefont {Gulka}}, \bibinfo {author}
  {\bibfnamefont {V.}~\bibnamefont {Iv\'ady}}, \bibinfo {author} {\bibfnamefont
  {F.}~\bibnamefont {Jelezko}}, \bibinfo {author} {\bibfnamefont
  {A.}~\bibnamefont {Gali}}, \bibinfo {author} {\bibfnamefont {M.}~\bibnamefont
  {Nesladek}},\ and\ \bibinfo {author} {\bibfnamefont {M.}~\bibnamefont
  {Trupke}},\ }\bibfield  {title} {\bibinfo {title} {Exploiting ionization
  dynamics in the nitrogen vacancy center for rapid, high-contrast spin, and
  charge state initialization},\ }\href
  {https://doi.org/10.1103/PhysRevResearch.5.013014} {\bibfield  {journal}
  {\bibinfo  {journal} {Phys. Rev. Res.}\ }\textbf {\bibinfo {volume} {5}},\
  \bibinfo {pages} {013014} (\bibinfo {year} {2023})}\BibitemShut {NoStop}%
\bibitem [{\citenamefont {Mims}(1972)}]{Mims1972}%
  \BibitemOpen
  \bibfield  {author} {\bibinfo {author} {\bibfnamefont {W.~B.}\ \bibnamefont
  {Mims}},\ }\bibfield  {title} {\bibinfo {title} {Envelope modulation in
  spin-echo experiments},\ }\href {https://doi.org/10.1103/PhysRevB.5.2409}
  {\bibfield  {journal} {\bibinfo  {journal} {Phys. Rev. B}\ }\textbf {\bibinfo
  {volume} {5}},\ \bibinfo {pages} {2409} (\bibinfo {year} {1972})}\BibitemShut
  {NoStop}%
\bibitem [{\citenamefont {Van't~Hof}\ and\ \citenamefont
  {Schmidt}(1979)}]{Van1979}%
  \BibitemOpen
  \bibfield  {author} {\bibinfo {author} {\bibfnamefont {C.}~\bibnamefont
  {Van't~Hof}}\ and\ \bibinfo {author} {\bibfnamefont {J.}~\bibnamefont
  {Schmidt}},\ }\bibfield  {title} {\bibinfo {title} {The effect of spectral
  diffusion on the phase coherence of phosphorescent triplet spins},\ }\href
  {https://doi.org/10.1080/00268977900101681} {\bibfield  {journal} {\bibinfo
  {journal} {Molecular Physics}\ }\textbf {\bibinfo {volume} {38}},\ \bibinfo
  {pages} {309} (\bibinfo {year} {1979})}\BibitemShut {NoStop}%
\bibitem [{\citenamefont {Janes}\ and\ \citenamefont
  {Brenner}(1983)}]{Janes1983}%
  \BibitemOpen
  \bibfield  {author} {\bibinfo {author} {\bibfnamefont {S.}~\bibnamefont
  {Janes}}\ and\ \bibinfo {author} {\bibfnamefont {H.}~\bibnamefont
  {Brenner}},\ }\bibfield  {title} {\bibinfo {title} {Triplet spin dephasing in
  impurity-induced traps in molecular crystals: detuning of proximate nuclear
  spins},\ }\href
  {https://doi.org/https://doi.org/10.1016/0009-2614(83)80803-3} {\bibfield
  {journal} {\bibinfo  {journal} {Chemical Physics Letters}\ }\textbf {\bibinfo
  {volume} {95}},\ \bibinfo {pages} {23} (\bibinfo {year} {1983})}\BibitemShut
  {NoStop}%
\bibitem [{\citenamefont {Weis}\ \emph {et~al.}(1998)\citenamefont {Weis},
  \citenamefont {M{\"o}bius},\ and\ \citenamefont {Prisner}}]{Weis1998}%
  \BibitemOpen
  \bibfield  {author} {\bibinfo {author} {\bibfnamefont {V.}~\bibnamefont
  {Weis}}, \bibinfo {author} {\bibfnamefont {K.}~\bibnamefont {M{\"o}bius}},\
  and\ \bibinfo {author} {\bibfnamefont {T.}~\bibnamefont {Prisner}},\
  }\bibfield  {title} {\bibinfo {title} {Optically detected electron spin echo
  envelope modulation on a photoexcited triplet state in zero magnetic
  field—a comparison between the zero-field and high-field limits},\ }\href
  {https://doi.org/https://doi.org/10.1006/jmre.1997.1340} {\bibfield
  {journal} {\bibinfo  {journal} {Journal of Magnetic Resonance}\ }\textbf
  {\bibinfo {volume} {131}},\ \bibinfo {pages} {17} (\bibinfo {year}
  {1998})}\BibitemShut {NoStop}%
\bibitem [{\citenamefont {Zaiser}\ \emph {et~al.}(2016)\citenamefont {Zaiser},
  \citenamefont {Rendler}, \citenamefont {Jakobi}, \citenamefont {Wolf},
  \citenamefont {Lee}, \citenamefont {Wagner}, \citenamefont {Bergholm},
  \citenamefont {Schulte-Herbr{\"u}ggen}, \citenamefont {Neumann},\ and\
  \citenamefont {Wrachtrup}}]{zaiser2016enhancing}%
  \BibitemOpen
  \bibfield  {author} {\bibinfo {author} {\bibfnamefont {S.}~\bibnamefont
  {Zaiser}}, \bibinfo {author} {\bibfnamefont {T.}~\bibnamefont {Rendler}},
  \bibinfo {author} {\bibfnamefont {I.}~\bibnamefont {Jakobi}}, \bibinfo
  {author} {\bibfnamefont {T.}~\bibnamefont {Wolf}}, \bibinfo {author}
  {\bibfnamefont {S.-Y.}\ \bibnamefont {Lee}}, \bibinfo {author} {\bibfnamefont
  {S.}~\bibnamefont {Wagner}}, \bibinfo {author} {\bibfnamefont
  {V.}~\bibnamefont {Bergholm}}, \bibinfo {author} {\bibfnamefont
  {T.}~\bibnamefont {Schulte-Herbr{\"u}ggen}}, \bibinfo {author} {\bibfnamefont
  {P.}~\bibnamefont {Neumann}},\ and\ \bibinfo {author} {\bibfnamefont
  {J.}~\bibnamefont {Wrachtrup}},\ }\bibfield  {title} {\bibinfo {title}
  {Enhancing quantum sensing sensitivity by a quantum memory},\ }\href
  {https://doi.org/10.1038/ncomms12279} {\bibfield  {journal} {\bibinfo
  {journal} {Nature communications}\ }\textbf {\bibinfo {volume} {7}},\
  \bibinfo {pages} {12279} (\bibinfo {year} {2016})}\BibitemShut {NoStop}%
\bibitem [{\citenamefont {Arunkumar}\ \emph {et~al.}(2023)\citenamefont
  {Arunkumar}, \citenamefont {Olsson}, \citenamefont {Oon}, \citenamefont
  {Hart}, \citenamefont {Bucher}, \citenamefont {Glenn}, \citenamefont {Lukin},
  \citenamefont {Park}, \citenamefont {Ham},\ and\ \citenamefont
  {Walsworth}}]{Arunkumar2023}%
  \BibitemOpen
  \bibfield  {author} {\bibinfo {author} {\bibfnamefont {N.}~\bibnamefont
  {Arunkumar}}, \bibinfo {author} {\bibfnamefont {K.~S.}\ \bibnamefont
  {Olsson}}, \bibinfo {author} {\bibfnamefont {J.~T.}\ \bibnamefont {Oon}},
  \bibinfo {author} {\bibfnamefont {C.~A.}\ \bibnamefont {Hart}}, \bibinfo
  {author} {\bibfnamefont {D.~B.}\ \bibnamefont {Bucher}}, \bibinfo {author}
  {\bibfnamefont {D.~R.}\ \bibnamefont {Glenn}}, \bibinfo {author}
  {\bibfnamefont {M.~D.}\ \bibnamefont {Lukin}}, \bibinfo {author}
  {\bibfnamefont {H.}~\bibnamefont {Park}}, \bibinfo {author} {\bibfnamefont
  {D.}~\bibnamefont {Ham}},\ and\ \bibinfo {author} {\bibfnamefont {R.~L.}\
  \bibnamefont {Walsworth}},\ }\bibfield  {title} {\bibinfo {title} {Quantum
  logic enhanced sensing in solid-state spin ensembles},\ }\href
  {https://doi.org/10.1103/PhysRevLett.131.100801} {\bibfield  {journal}
  {\bibinfo  {journal} {Phys. Rev. Lett.}\ }\textbf {\bibinfo {volume} {131}},\
  \bibinfo {pages} {100801} (\bibinfo {year} {2023})}\BibitemShut {NoStop}%
\bibitem [{\citenamefont {Wu}\ \emph {et~al.}(2019)\citenamefont {Wu},
  \citenamefont {Ng}, \citenamefont {Mirkhanov}, \citenamefont {Amirzhan},
  \citenamefont {Nitnara},\ and\ \citenamefont {Oxborrow}}]{Wu2019}%
  \BibitemOpen
  \bibfield  {author} {\bibinfo {author} {\bibfnamefont {H.}~\bibnamefont
  {Wu}}, \bibinfo {author} {\bibfnamefont {W.}~\bibnamefont {Ng}}, \bibinfo
  {author} {\bibfnamefont {S.}~\bibnamefont {Mirkhanov}}, \bibinfo {author}
  {\bibfnamefont {A.}~\bibnamefont {Amirzhan}}, \bibinfo {author}
  {\bibfnamefont {S.}~\bibnamefont {Nitnara}},\ and\ \bibinfo {author}
  {\bibfnamefont {M.}~\bibnamefont {Oxborrow}},\ }\bibfield  {title} {\bibinfo
  {title} {Unraveling the room-temperature spin dynamics of photoexcited
  pentacene in its lowest triplet state at zero field},\ }\href
  {https://doi.org/10.1021/acs.jpcc.9b08439} {\bibfield  {journal} {\bibinfo
  {journal} {The Journal of Physical Chemistry C}\ }\textbf {\bibinfo {volume}
  {123}},\ \bibinfo {pages} {24275} (\bibinfo {year} {2019})}\BibitemShut
  {NoStop}%
\bibitem [{\citenamefont {Attwood}\ \emph {et~al.}(2023)\citenamefont
  {Attwood}, \citenamefont {Xu}, \citenamefont {Newns}, \citenamefont {Meng},
  \citenamefont {Ingle}, \citenamefont {Wu}, \citenamefont {Chen},
  \citenamefont {Xu}, \citenamefont {Ng}, \citenamefont {Abiola}, \citenamefont
  {Stavros},\ and\ \citenamefont {Oxborrow}}]{Attwood2023}%
  \BibitemOpen
  \bibfield  {author} {\bibinfo {author} {\bibfnamefont {M.}~\bibnamefont
  {Attwood}}, \bibinfo {author} {\bibfnamefont {X.}~\bibnamefont {Xu}},
  \bibinfo {author} {\bibfnamefont {M.}~\bibnamefont {Newns}}, \bibinfo
  {author} {\bibfnamefont {Z.}~\bibnamefont {Meng}}, \bibinfo {author}
  {\bibfnamefont {R.~A.}\ \bibnamefont {Ingle}}, \bibinfo {author}
  {\bibfnamefont {H.}~\bibnamefont {Wu}}, \bibinfo {author} {\bibfnamefont
  {X.}~\bibnamefont {Chen}}, \bibinfo {author} {\bibfnamefont {W.}~\bibnamefont
  {Xu}}, \bibinfo {author} {\bibfnamefont {W.}~\bibnamefont {Ng}}, \bibinfo
  {author} {\bibfnamefont {T.~T.}\ \bibnamefont {Abiola}}, \bibinfo {author}
  {\bibfnamefont {V.~G.}\ \bibnamefont {Stavros}},\ and\ \bibinfo {author}
  {\bibfnamefont {M.}~\bibnamefont {Oxborrow}},\ }\bibfield  {title} {\bibinfo
  {title} {N-heteroacenes as an organic gain medium for room-temperature
  masers},\ }\href {https://doi.org/10.1021/acs.chemmater.3c00640} {\bibfield
  {journal} {\bibinfo  {journal} {Chemistry of Materials}\ }\textbf {\bibinfo
  {volume} {35}},\ \bibinfo {pages} {4498} (\bibinfo {year}
  {2023})}\BibitemShut {NoStop}%
\bibitem [{\citenamefont {Wu}\ \emph {et~al.}(2020)\citenamefont {Wu},
  \citenamefont {Xie}, \citenamefont {Ng}, \citenamefont {Mehanna},
  \citenamefont {Li}, \citenamefont {Attwood},\ and\ \citenamefont
  {Oxborrow}}]{Wu2020}%
  \BibitemOpen
  \bibfield  {author} {\bibinfo {author} {\bibfnamefont {H.}~\bibnamefont
  {Wu}}, \bibinfo {author} {\bibfnamefont {X.}~\bibnamefont {Xie}}, \bibinfo
  {author} {\bibfnamefont {W.}~\bibnamefont {Ng}}, \bibinfo {author}
  {\bibfnamefont {S.}~\bibnamefont {Mehanna}}, \bibinfo {author} {\bibfnamefont
  {Y.}~\bibnamefont {Li}}, \bibinfo {author} {\bibfnamefont {M.}~\bibnamefont
  {Attwood}},\ and\ \bibinfo {author} {\bibfnamefont {M.}~\bibnamefont
  {Oxborrow}},\ }\bibfield  {title} {\bibinfo {title} {Room-temperature
  quasi-continuous-wave pentacene maser pumped by an invasive
  $\mathrm{Ce}:\mathrm{YAG}$ luminescent concentrator},\ }\href
  {https://doi.org/10.1103/PhysRevApplied.14.064017} {\bibfield  {journal}
  {\bibinfo  {journal} {Phys. Rev. Appl.}\ }\textbf {\bibinfo {volume} {14}},\
  \bibinfo {pages} {064017} (\bibinfo {year} {2020})}\BibitemShut {NoStop}%
\bibitem [{\citenamefont {Siebrand}(1970)}]{Siebrand1970}%
  \BibitemOpen
  \bibfield  {author} {\bibinfo {author} {\bibfnamefont {W.}~\bibnamefont
  {Siebrand}},\ }\bibfield  {title} {\bibinfo {title} {Mechanisms of
  intersystem crossing in aromatic hydrocarbons},\ }\href
  {https://doi.org/https://doi.org/10.1016/0009-2614(70)80215-9} {\bibfield
  {journal} {\bibinfo  {journal} {Chemical Physics Letters}\ }\textbf {\bibinfo
  {volume} {6}},\ \bibinfo {pages} {192} (\bibinfo {year} {1970})}\BibitemShut
  {NoStop}%
\bibitem [{\citenamefont {Metz}(1973)}]{Metz1973}%
  \BibitemOpen
  \bibfield  {author} {\bibinfo {author} {\bibfnamefont {F.}~\bibnamefont
  {Metz}},\ }\bibfield  {title} {\bibinfo {title} {Position-dependent deuterium
  effect on relative rate constants for isc processes in aromatic
  hydrocarbons},\ }\href
  {https://doi.org/https://doi.org/10.1016/0009-2614(73)80567-6} {\bibfield
  {journal} {\bibinfo  {journal} {Chemical Physics Letters}\ }\textbf {\bibinfo
  {volume} {22}},\ \bibinfo {pages} {186} (\bibinfo {year} {1973})}\BibitemShut
  {NoStop}%
\bibitem [{\citenamefont {{Clarke}}\ and\ \citenamefont
  {{Frank}}(1976)}]{Clarke1976}%
  \BibitemOpen
  \bibfield  {author} {\bibinfo {author} {\bibfnamefont {R.~H.}\ \bibnamefont
  {{Clarke}}}\ and\ \bibinfo {author} {\bibfnamefont {H.~A.}\ \bibnamefont
  {{Frank}}},\ }\bibfield  {title} {\bibinfo {title} {{Triplet state
  radiationless transitions in polycyclic hydrocarbons}},\ }\href
  {https://doi.org/10.1063/1.432781} {\bibfield  {journal} {\bibinfo  {journal}
  {\jcp}\ }\textbf {\bibinfo {volume} {65}},\ \bibinfo {pages} {39} (\bibinfo
  {year} {1976})}\BibitemShut {NoStop}%
\bibitem [{\citenamefont {El-Sayed}(1963)}]{el1963spin}%
  \BibitemOpen
  \bibfield  {author} {\bibinfo {author} {\bibfnamefont {M.}~\bibnamefont
  {El-Sayed}},\ }\bibfield  {title} {\bibinfo {title} {Spin—orbit coupling
  and the radiationless processes in nitrogen heterocyclics},\ }\href
  {https://doi.org/https://doi.org/10.1063/1.1733610} {\bibfield  {journal}
  {\bibinfo  {journal} {The Journal of Chemical Physics}\ }\textbf {\bibinfo
  {volume} {38}},\ \bibinfo {pages} {2834} (\bibinfo {year}
  {1963})}\BibitemShut {NoStop}%
\bibitem [{\citenamefont {Marian}(2021)}]{marian2021understanding}%
  \BibitemOpen
  \bibfield  {author} {\bibinfo {author} {\bibfnamefont {C.~M.}\ \bibnamefont
  {Marian}},\ }\bibfield  {title} {\bibinfo {title} {Understanding and
  controlling intersystem crossing in molecules},\ }\href
  {https://doi.org/https://doi.org/10.1146/annurev-physchem-061020-053433}
  {\bibfield  {journal} {\bibinfo  {journal} {Annual review of physical
  chemistry}\ }\textbf {\bibinfo {volume} {72}},\ \bibinfo {pages} {617}
  (\bibinfo {year} {2021})}\BibitemShut {NoStop}%
\bibitem [{\citenamefont {Gromer}\ \emph {et~al.}(1972)\citenamefont {Gromer},
  \citenamefont {Sixl},\ and\ \citenamefont {Wolf}}]{Gromer1972}%
  \BibitemOpen
  \bibfield  {author} {\bibinfo {author} {\bibfnamefont {J.}~\bibnamefont
  {Gromer}}, \bibinfo {author} {\bibfnamefont {H.}~\bibnamefont {Sixl}},\ and\
  \bibinfo {author} {\bibfnamefont {H.}~\bibnamefont {Wolf}},\ }\bibfield
  {title} {\bibinfo {title} {{Optical electron spin polarisation: influence of
  deuteration and energy transfer}},\ }\href
  {https://doi.org/10.1016/0009-2614(72)80011-3} {\bibfield  {journal}
  {\bibinfo  {journal} {Chemical Physics Letters}\ }\textbf {\bibinfo {volume}
  {12}},\ \bibinfo {pages} {574} (\bibinfo {year} {1972})}\BibitemShut
  {NoStop}%
\bibitem [{\citenamefont {D.A.~Antheunis}\ and\ \citenamefont {van~der
  Waals}(1974)}]{Antheunis1974}%
  \BibitemOpen
  \bibfield  {author} {\bibinfo {author} {\bibfnamefont {J.~S.}\ \bibnamefont
  {D.A.~Antheunis}}\ and\ \bibinfo {author} {\bibfnamefont {J.}~\bibnamefont
  {van~der Waals}},\ }\bibfield  {title} {\bibinfo {title} {Spin-forbidden
  radiationless processes in isoelectronic molecules: Anthracene, acridine and
  phenazine},\ }\href {https://doi.org/10.1080/00268977400101291} {\bibfield
  {journal} {\bibinfo  {journal} {Molecular Physics}\ }\textbf {\bibinfo
  {volume} {27}},\ \bibinfo {pages} {1521} (\bibinfo {year}
  {1974})}\BibitemShut {NoStop}%
\bibitem [{\citenamefont {McGuinness}\ \emph {et~al.}(2011)\citenamefont
  {McGuinness}, \citenamefont {Yan}, \citenamefont {Stacey}, \citenamefont
  {Simpson}, \citenamefont {Hall}, \citenamefont {Maclaurin}, \citenamefont
  {Prawer}, \citenamefont {Mulvaney}, \citenamefont {Wrachtrup}, \citenamefont
  {Caruso}, \citenamefont {Scholten},\ and\ \citenamefont
  {Hollenberg}}]{McGuinness2011}%
  \BibitemOpen
  \bibfield  {author} {\bibinfo {author} {\bibfnamefont {L.~P.}\ \bibnamefont
  {McGuinness}}, \bibinfo {author} {\bibfnamefont {Y.}~\bibnamefont {Yan}},
  \bibinfo {author} {\bibfnamefont {A.}~\bibnamefont {Stacey}}, \bibinfo
  {author} {\bibfnamefont {D.~A.}\ \bibnamefont {Simpson}}, \bibinfo {author}
  {\bibfnamefont {L.~T.}\ \bibnamefont {Hall}}, \bibinfo {author}
  {\bibfnamefont {D.}~\bibnamefont {Maclaurin}}, \bibinfo {author}
  {\bibfnamefont {S.}~\bibnamefont {Prawer}}, \bibinfo {author} {\bibfnamefont
  {P.}~\bibnamefont {Mulvaney}}, \bibinfo {author} {\bibfnamefont
  {J.}~\bibnamefont {Wrachtrup}}, \bibinfo {author} {\bibfnamefont
  {F.}~\bibnamefont {Caruso}}, \bibinfo {author} {\bibfnamefont {R.~E.}\
  \bibnamefont {Scholten}},\ and\ \bibinfo {author} {\bibfnamefont {L.~C.~L.}\
  \bibnamefont {Hollenberg}},\ }\bibfield  {title} {\bibinfo {title} {Quantum
  measurement and orientation tracking of fluorescent nanodiamonds inside
  living cells},\ }\href {https://doi.org/10.1038/nnano.2011.64} {\bibfield
  {journal} {\bibinfo  {journal} {Nature Nanotechnology}\ }\textbf {\bibinfo
  {volume} {6}},\ \bibinfo {pages} {358} (\bibinfo {year} {2011})}\BibitemShut
  {NoStop}%
\bibitem [{\citenamefont {Kucsko}\ \emph {et~al.}(2013)\citenamefont {Kucsko},
  \citenamefont {Maurer}, \citenamefont {Yao}, \citenamefont {Kubo},
  \citenamefont {Noh}, \citenamefont {Lo}, \citenamefont {Park},\ and\
  \citenamefont {Lukin}}]{kucsko2013nanometre}%
  \BibitemOpen
  \bibfield  {author} {\bibinfo {author} {\bibfnamefont {G.}~\bibnamefont
  {Kucsko}}, \bibinfo {author} {\bibfnamefont {P.~C.}\ \bibnamefont {Maurer}},
  \bibinfo {author} {\bibfnamefont {N.~Y.}\ \bibnamefont {Yao}}, \bibinfo
  {author} {\bibfnamefont {M.}~\bibnamefont {Kubo}}, \bibinfo {author}
  {\bibfnamefont {H.~J.}\ \bibnamefont {Noh}}, \bibinfo {author} {\bibfnamefont
  {P.~K.}\ \bibnamefont {Lo}}, \bibinfo {author} {\bibfnamefont
  {H.}~\bibnamefont {Park}},\ and\ \bibinfo {author} {\bibfnamefont {M.~D.}\
  \bibnamefont {Lukin}},\ }\bibfield  {title} {\bibinfo {title}
  {Nanometre-scale thermometry in a living cell},\ }\href
  {https://doi.org/https://doi.org/10.1038/nature12373} {\bibfield  {journal}
  {\bibinfo  {journal} {Nature}\ }\textbf {\bibinfo {volume} {500}},\ \bibinfo
  {pages} {54} (\bibinfo {year} {2013})}\BibitemShut {NoStop}%
\bibitem [{\citenamefont {Pazzagli}\ \emph {et~al.}(2018)\citenamefont
  {Pazzagli}, \citenamefont {Lombardi}, \citenamefont {Martella}, \citenamefont
  {Colautti}, \citenamefont {Tiribilli}, \citenamefont {Cataliotti},\ and\
  \citenamefont {Toninelli}}]{Pazzagli2018}%
  \BibitemOpen
  \bibfield  {author} {\bibinfo {author} {\bibfnamefont {S.}~\bibnamefont
  {Pazzagli}}, \bibinfo {author} {\bibfnamefont {P.}~\bibnamefont {Lombardi}},
  \bibinfo {author} {\bibfnamefont {D.}~\bibnamefont {Martella}}, \bibinfo
  {author} {\bibfnamefont {M.}~\bibnamefont {Colautti}}, \bibinfo {author}
  {\bibfnamefont {B.}~\bibnamefont {Tiribilli}}, \bibinfo {author}
  {\bibfnamefont {F.~S.}\ \bibnamefont {Cataliotti}},\ and\ \bibinfo {author}
  {\bibfnamefont {C.}~\bibnamefont {Toninelli}},\ }\bibfield  {title} {\bibinfo
  {title} {Self-assembled nanocrystals of polycyclic aromatic hydrocarbons show
  photostable single-photon emission},\ }\href
  {https://doi.org/10.1021/acsnano.7b08810} {\bibfield  {journal} {\bibinfo
  {journal} {ACS Nano}\ }\textbf {\bibinfo {volume} {12}},\ \bibinfo {pages}
  {4295} (\bibinfo {year} {2018})}\BibitemShut {NoStop}%
\bibitem [{\citenamefont {Toninelli}\ \emph {et~al.}(2021)\citenamefont
  {Toninelli}, \citenamefont {Gerhardt}, \citenamefont {Clark}, \citenamefont
  {Reserbat-Plantey}, \citenamefont {G{\"o}tzinger}, \citenamefont
  {Ristanovi{\'c}}, \citenamefont {Colautti}, \citenamefont {Lombardi},
  \citenamefont {Major}, \citenamefont {Deperasi{\'n}ska} \emph
  {et~al.}}]{toninelli2021single}%
  \BibitemOpen
  \bibfield  {author} {\bibinfo {author} {\bibfnamefont {C.}~\bibnamefont
  {Toninelli}}, \bibinfo {author} {\bibfnamefont {I.}~\bibnamefont {Gerhardt}},
  \bibinfo {author} {\bibfnamefont {A.}~\bibnamefont {Clark}}, \bibinfo
  {author} {\bibfnamefont {A.}~\bibnamefont {Reserbat-Plantey}}, \bibinfo
  {author} {\bibfnamefont {S.}~\bibnamefont {G{\"o}tzinger}}, \bibinfo {author}
  {\bibfnamefont {Z.}~\bibnamefont {Ristanovi{\'c}}}, \bibinfo {author}
  {\bibfnamefont {M.}~\bibnamefont {Colautti}}, \bibinfo {author}
  {\bibfnamefont {P.}~\bibnamefont {Lombardi}}, \bibinfo {author}
  {\bibfnamefont {K.}~\bibnamefont {Major}}, \bibinfo {author} {\bibfnamefont
  {I.}~\bibnamefont {Deperasi{\'n}ska}}, \emph {et~al.},\ }\bibfield  {title}
  {\bibinfo {title} {Single organic molecules for photonic quantum
  technologies},\ }\href
  {https://doi.org/https://doi.org/10.1038/s41563-021-00987-4} {\bibfield
  {journal} {\bibinfo  {journal} {Nature Materials}\ }\textbf {\bibinfo
  {volume} {20}},\ \bibinfo {pages} {1615} (\bibinfo {year}
  {2021})}\BibitemShut {NoStop}%
\bibitem [{\citenamefont {Kasai}\ \emph {et~al.}(1992)\citenamefont {Kasai},
  \citenamefont {Nalwa}, \citenamefont {Oikawa}, \citenamefont {Okada},
  \citenamefont {Matsuda}, \citenamefont {Minami}, \citenamefont {Kakuta},
  \citenamefont {Ono}, \citenamefont {Mukoh},\ and\ \citenamefont
  {Hachiro~Nakanishi}}]{Kasai1992}%
  \BibitemOpen
  \bibfield  {author} {\bibinfo {author} {\bibfnamefont {H.}~\bibnamefont
  {Kasai}}, \bibinfo {author} {\bibfnamefont {H.~S.}\ \bibnamefont {Nalwa}},
  \bibinfo {author} {\bibfnamefont {H.}~\bibnamefont {Oikawa}}, \bibinfo
  {author} {\bibfnamefont {S.}~\bibnamefont {Okada}}, \bibinfo {author}
  {\bibfnamefont {H.}~\bibnamefont {Matsuda}}, \bibinfo {author} {\bibfnamefont
  {N.}~\bibnamefont {Minami}}, \bibinfo {author} {\bibfnamefont
  {A.}~\bibnamefont {Kakuta}}, \bibinfo {author} {\bibfnamefont
  {K.}~\bibnamefont {Ono}}, \bibinfo {author} {\bibfnamefont {A.}~\bibnamefont
  {Mukoh}},\ and\ \bibinfo {author} {\bibfnamefont {H.~N.}\ \bibnamefont
  {Hachiro~Nakanishi}},\ }\bibfield  {title} {\bibinfo {title} {A novel
  preparation method of organic microcrystals},\ }\href
  {https://doi.org/10.1143/JJAP.31.L1132} {\bibfield  {journal} {\bibinfo
  {journal} {Japanese Journal of Applied Physics}\ }\textbf {\bibinfo {volume}
  {31}},\ \bibinfo {pages} {L1132} (\bibinfo {year} {1992})}\BibitemShut
  {NoStop}%
\bibitem [{\citenamefont {Kang}\ \emph {et~al.}(2004)\citenamefont {Kang},
  \citenamefont {Chen}, \citenamefont {Hao}, \citenamefont {Zhu}, \citenamefont
  {Hu},\ and\ \citenamefont {Chen}}]{kang2004novel}%
  \BibitemOpen
  \bibfield  {author} {\bibinfo {author} {\bibfnamefont {P.}~\bibnamefont
  {Kang}}, \bibinfo {author} {\bibfnamefont {C.}~\bibnamefont {Chen}}, \bibinfo
  {author} {\bibfnamefont {L.}~\bibnamefont {Hao}}, \bibinfo {author}
  {\bibfnamefont {C.}~\bibnamefont {Zhu}}, \bibinfo {author} {\bibfnamefont
  {Y.}~\bibnamefont {Hu}},\ and\ \bibinfo {author} {\bibfnamefont
  {Z.}~\bibnamefont {Chen}},\ }\bibfield  {title} {\bibinfo {title} {A novel
  sonication route to prepare anthracene nanoparticles},\ }\href
  {https://doi.org/10.1016/j.materresbull.2003.12.013} {\bibfield  {journal}
  {\bibinfo  {journal} {Materials Research Bulletin}\ }\textbf {\bibinfo
  {volume} {39}},\ \bibinfo {pages} {545} (\bibinfo {year} {2004})}\BibitemShut
  {NoStop}%
\bibitem [{\citenamefont {Baba}\ \emph {et~al.}(2011)\citenamefont {Baba},
  \citenamefont {Kasai}, \citenamefont {Nishida},\ and\ \citenamefont
  {Nakanishi}}]{Baba2011}%
  \BibitemOpen
  \bibfield  {author} {\bibinfo {author} {\bibfnamefont {K.}~\bibnamefont
  {Baba}}, \bibinfo {author} {\bibfnamefont {H.}~\bibnamefont {Kasai}},
  \bibinfo {author} {\bibfnamefont {K.}~\bibnamefont {Nishida}},\ and\ \bibinfo
  {author} {\bibfnamefont {H.}~\bibnamefont {Nakanishi}},\ }\bibfield  {title}
  {\bibinfo {title} {Functional organic nanocrystals},\ }\href
  {https://doi.org/DOI: 10.5772/16948} {\bibfield  {journal} {\bibinfo
  {journal} {Nanocrystals, ed. Y. Masuda (IntechOpen, 2011) Ch}\ }\textbf
  {\bibinfo {volume} {15}},\ \bibinfo {pages} {397} (\bibinfo {year}
  {2011})}\BibitemShut {NoStop}%
\bibitem [{\citenamefont {Oxborrow}\ \emph {et~al.}(2012)\citenamefont
  {Oxborrow}, \citenamefont {Breeze},\ and\ \citenamefont
  {Alford}}]{Oxborrow2012}%
  \BibitemOpen
  \bibfield  {author} {\bibinfo {author} {\bibfnamefont {M.}~\bibnamefont
  {Oxborrow}}, \bibinfo {author} {\bibfnamefont {J.~D.}\ \bibnamefont
  {Breeze}},\ and\ \bibinfo {author} {\bibfnamefont {N.~M.}\ \bibnamefont
  {Alford}},\ }\bibfield  {title} {\bibinfo {title} {Room-temperature
  solid-state maser},\ }\href {https://doi.org/10.1038/nature11339} {\bibfield
  {journal} {\bibinfo  {journal} {Nature}\ }\textbf {\bibinfo {volume} {488}},\
  \bibinfo {pages} {353} (\bibinfo {year} {2012})}\BibitemShut {NoStop}%
\bibitem [{\citenamefont {Filidou}\ \emph {et~al.}(2012)\citenamefont
  {Filidou}, \citenamefont {Simmons}, \citenamefont {Karlen}, \citenamefont
  {Giustino}, \citenamefont {Anderson},\ and\ \citenamefont
  {Morton}}]{filidou2012ultrafast}%
  \BibitemOpen
  \bibfield  {author} {\bibinfo {author} {\bibfnamefont {V.}~\bibnamefont
  {Filidou}}, \bibinfo {author} {\bibfnamefont {S.}~\bibnamefont {Simmons}},
  \bibinfo {author} {\bibfnamefont {S.~D.}\ \bibnamefont {Karlen}}, \bibinfo
  {author} {\bibfnamefont {F.}~\bibnamefont {Giustino}}, \bibinfo {author}
  {\bibfnamefont {H.~L.}\ \bibnamefont {Anderson}},\ and\ \bibinfo {author}
  {\bibfnamefont {J.~J.}\ \bibnamefont {Morton}},\ }\bibfield  {title}
  {\bibinfo {title} {Ultrafast entangling gates between nuclear spins using
  photoexcited triplet states},\ }\href
  {https://doi.org/https://doi.org/10.1038/nphys2353} {\bibfield  {journal}
  {\bibinfo  {journal} {Nature Physics}\ }\textbf {\bibinfo {volume} {8}},\
  \bibinfo {pages} {596} (\bibinfo {year} {2012})}\BibitemShut {NoStop}%
\bibitem [{\citenamefont {Lee}\ \emph {et~al.}(2013)\citenamefont {Lee},
  \citenamefont {Widmann}, \citenamefont {Rendler}, \citenamefont {Doherty},
  \citenamefont {Babinec}, \citenamefont {Yang}, \citenamefont {Eyer},
  \citenamefont {Siyushev}, \citenamefont {Hausmann}, \citenamefont {Loncar},
  \citenamefont {Bodrog}, \citenamefont {Gali}, \citenamefont {Manson},
  \citenamefont {Fedder},\ and\ \citenamefont {Wrachtrup}}]{Lee2013_ancilla}%
  \BibitemOpen
  \bibfield  {author} {\bibinfo {author} {\bibfnamefont {S.-Y.}\ \bibnamefont
  {Lee}}, \bibinfo {author} {\bibfnamefont {M.}~\bibnamefont {Widmann}},
  \bibinfo {author} {\bibfnamefont {T.}~\bibnamefont {Rendler}}, \bibinfo
  {author} {\bibfnamefont {M.~W.}\ \bibnamefont {Doherty}}, \bibinfo {author}
  {\bibfnamefont {T.~M.}\ \bibnamefont {Babinec}}, \bibinfo {author}
  {\bibfnamefont {S.}~\bibnamefont {Yang}}, \bibinfo {author} {\bibfnamefont
  {M.}~\bibnamefont {Eyer}}, \bibinfo {author} {\bibfnamefont {P.}~\bibnamefont
  {Siyushev}}, \bibinfo {author} {\bibfnamefont {B.~J.~M.}\ \bibnamefont
  {Hausmann}}, \bibinfo {author} {\bibfnamefont {M.}~\bibnamefont {Loncar}},
  \bibinfo {author} {\bibfnamefont {Z.}~\bibnamefont {Bodrog}}, \bibinfo
  {author} {\bibfnamefont {A.}~\bibnamefont {Gali}}, \bibinfo {author}
  {\bibfnamefont {N.~B.}\ \bibnamefont {Manson}}, \bibinfo {author}
  {\bibfnamefont {H.}~\bibnamefont {Fedder}},\ and\ \bibinfo {author}
  {\bibfnamefont {J.}~\bibnamefont {Wrachtrup}},\ }\bibfield  {title} {\bibinfo
  {title} {Readout and control of a single nuclear spin with a metastable
  electron spin ancilla},\ }\href {https://doi.org/10.1038/nnano.2013.104}
  {\bibfield  {journal} {\bibinfo  {journal} {Nature Nanotechnology}\ }\textbf
  {\bibinfo {volume} {8}},\ \bibinfo {pages} {487} (\bibinfo {year}
  {2013})}\BibitemShut {NoStop}%
\end{thebibliography}
\end{document}